\documentclass[twocolumn,showpacs,amsmath,amssymb]{revtex4-1}
\usepackage{hyperref}
\input{epsf}

\usepackage{graphicx}
\usepackage{array}
\usepackage{longtable}
\usepackage{rotating,booktabs}
\usepackage{booktabs,threeparttable}
\usepackage{bm}
\usepackage[left]{lineno}
\usepackage{graphicx}
\usepackage{epstopdf}

\usepackage{color}

\begin{document}
\title{Tune-out wavelengths of the hyperfine components of the ground level of $^{133}$Cs atoms}

\author{Jun Jiang}
\email {phyjiang@yeah.net}
\author{Xian-Jun Li}
\author{Xia Wang}
\author{Chen-Zhong Dong}
\author{Z.~W.~Wu}

\affiliation{$^1$Key Laboratory of Atomic and Molecular Physics and Functional Materials of Gansu Province, College of Physics and Electronic Engineering, Northwest Normal University, Lanzhou 730070, P. R. China}

\date{\today}

\begin{abstract}
The static and dynamic electric-dipole polarizabilities and tune-out wavelengths 
for the ground state of Cs atoms are calculated by using 
a semiempirical relativistic 
configuration interaction plus core polarization approach. 
By considering the hyperfine splittings, the static and dynamic polarizabilities, 
hyperfine Stark shifts, 
and tune-out wavelengths of the hyperfine components of the ground level of $^{133}$Cs atoms are determined. 
It is found that the hyperfine shifts of the first primary 
tune-out wavelengths are about $-$0.0135 nm and 0.0106~nm, 
which are very close to the hyperfine interaction energies. 
Additionally, the contribution of the tensor 
polarizability to the tune-out wavelengths is determined to be about $10^{-5} \sim 10^{-6}$ nm.
\end{abstract}

 \maketitle

\section{Introduction}

With the rapid development of laser cooling and trapping technologies,
high-precision measurements of atomic spectra were remarkably developed in the past two decades.
In an optical lattice, atoms could be trapped by a dipole force proportional
to the dynamic polarizability~\cite{Grimm2000}.
At certain laser wavelengths, however, the dynamic polarizability becomes null and, accordingly, the dipole force between atoms and laser field vanishes as well. These laser wavelengths are termed as ``tune-out wavelengths" by LeBlanc and Thywissen~\cite{LeBlanc2007}. The tune-out wavelengths are very useful for sympathetic cooling in two-species mixtures of alkali-metal atoms such as Li-Cs~\cite{LeBlanc2007, Luan2015}, K-Rb~\cite{LeBlanc2007, Chamakhi2015, Catani2009}, Rb-Cs~\cite{LeBlanc2007}, K-Cs~\cite{LeBlanc2007} and $^{39}$K-$^{40}$K~\cite{LeBlanc2007} mixtures.

High-precision measurements of tune-out wavelengths can be used to determine 
atomic parameters such as the ratios of line strengths or oscillator strengths~\cite{Herold2012,Safronova2015,Fallon2016}. 
For example, the longest tune-out wavelength of the ground state of K was measured 
with an uncertainty of 1.5 pm~\cite{Holmgren2012}, and the ratio of the $4s \rightarrow 4p_j$ line strengths 
was obtained with an accuracy of 0.0005.
The longest tune-out wavelength of the $5s_{1/2}, F=2$ state of  $^{87}$Rb was measured with an accuracy of about 
30 fm~\cite{Leonard2015, Leonard2017}, and the ratio of  the $5s_{1/2}-5p_j$ line strengths 
was determined with an accuracy of 0.007. The longest tune-out wavelength of the metastable state of 
He has been measured with an uncertainty of 1.8 pm~\cite{Henson2015}. 
This tune-out wavelength has been recommended to test the quantum 
electrodynamics (QED) effect~\cite{Henson2015, Mitroy2013}. 
Additionally, a tune-out wavelength of $^7$Li was measured with an uncertainty of 2.2 fm~\cite{Copenhaver2019}.

Cs atoms are used in a wide range of applications including atomic clocks, 
quantum information, and quantum communication~\cite{Demonstration1995, Jefferts2014, WangYang2015}.
These applications require an accurate understanding of atomic properties of Cs atoms, in particular,
electric-dipole (E1) matrix elements and polarizabilities. In order to obtain highly
accurate E1 transition matrix elements of Cs atoms,
high-precision measurements and calculations of its tune-out wavelengths are required.
Arora \emph{et al.}~\cite{Arora2011} calculated the tune-out wavelengths
of the $6s_{1/2}$ state of Cs atoms by using a relativistic all-order method. However, the effect of the hyperfine interactions on the tune-out wavelength was not considered.
In order to obtain high-precision tune-out wavelengths, the contribution of the hyperfine interactions
should be taken into account.
To the best of our knowledge, the tune-out wavelengths of Cs atoms are not experimentally measured yet.

In this paper, the static and dynamic polarizabilities as well as three tune-out wavelengths are calculated 
by using a semi-empirical method---the so-called relativistic configuration interaction plus core polarization (RCICP) method, 
which was developed based on a nonrelativistic configuration interaction method plus a core polarization model~\cite{Mitroy2003Semiempirical}.
By considering the hyperfine splittings, the E1 matrix elements between the hyperfine states, 
the static and dynamic polarizabilities and the tune-out wavelengths of the hyperfine components of the ground level of $^{133}$Cs atoms are determined.
Atomic units are used throughout the text unless specified otherwise.

\section{Theory and Calculations for the $6s_{1/2}$ ground state of Cs atoms}

\subsection{Energies of low-lying states}

The basic strategy of the theoretical model used is to partition a Cs atom into a core Cs$^{+}$ plus a valence electron. 
The first step involves a Dirac-Fock~(DF) calculation of the core. 
The orbitals of the core are expressed as linear combinations of $S$-spinors,
which can be treated as a relativistic generalization of the Slater-type orbitals~\cite{ Grant2007, grant1988, grant1989, grant1996}. 
The Hamiltonian of the valence electron is written as
\begin{eqnarray}
    {\bf \emph{H}} = c \bm{\alpha} \cdot \bm{p}+(\beta-1)c^2+V_{core}(\bm{r}),
 \end{eqnarray}
where $\bm{\alpha}$ and $\beta$ are $4\times4$ matrices of the Dirac operator, $\bm{p}$ is the momentum operator, 
and $c$ is the speed of light, $\bm{r}$ is the position vector of the valence electron. 
The effective interaction potential $V_{core}(\bm{r})$ can be expressed as
\begin {eqnarray}
 V_{core}(\bm{r})=-\frac{Z}{r}+V_{dir}(\bm{r})+V_{exc}(\bm {r})+V_{p}(\bm{r}).
\end{eqnarray}
Here, $Z$ is atomic number, $r$ is the distance of  the  valence electron with respect to the origin of the coordinates, 
the direct and exchange interactions $V_{dir}$ and $V_{exc}$ of the valence electron with the DF core are calculated exactly, 
while the $\ell$-dependent polarization potential $V_p$ is treated semiempirically as follows,
\begin{eqnarray}
	V_{p}(\bm{r})= -\sum_{k=1}^{2} \frac{\alpha_{\mathrm{core}}^{(k)}}{2r^{2(k+1)}}
	\sum_{\ell,j} g^2_{\ell,j}(r) | \ell j \rangle \langle \ell j|.
\end{eqnarray}
Here, $\ell$  is the orbital angular momentum quantum number, $j$ is 
the total angular momentum quantum number, the factor $\alpha_{\mathrm{core}}^{(k)}$ is 
the $k$th-order static polarizability of the core with $\alpha_{\mathrm{core}}^{(1)}$ = 15.8(1) a.u.~\cite{Lim2002}
for dipole polarizability and $\alpha_{\mathrm{core}}^{(2)}$ = 86.4 a.u.~\cite{Johnson1983} 
for quadrupole polarizability. $g_{\ell ,j}^{2}(r)=1-$exp$(-r^{6}/\rho_{\ell ,j}^{6})$ is 
a cutoff function designed to make the polarization potential finite at the origin. 
The cutoff parameters $\rho_{{\ell},j}$, as listed in Table~\ref{rho}, are 
tuned to reproduce the binding energies of the ground state and some low-lying excited states.
\begin{table}
\caption{\label{rho} Cutoff parameters $\rho_{\ell, j}$ of the polarization potential of the Cs$^{+}$ core.}
\begin{ruledtabular}
\begin{tabular}{lll}
$\ell$ & $j$ & $\rho_{\ell, j}$~(a.u.) \\ \hline
$s$ & 1/2 & 2.78449  \\
$p$ & 1/2 & 2.66646  \\
       & 3/2 & 2.68025 \\
$d$ & 3/2 & 3.19976 \\
       & 5/2 & 3.24245 \\
\end{tabular}
\end{ruledtabular}
\end{table}

The corresponding effective Hamiltonian is diagonalized in a large $L$-spinor basis~\cite{Grant2007,Grant2000}. 
$L$-spinors can be regarded as a relativistic generalization of 
the Laguerre-type orbitals that are often utilized in solving the Schr{\"o}dinger equation~\cite{Mitroy2003}. 
The present calculations used 50 positive energy and 50 negative energy $L$-spinors for each $(\ell,~j )$ symmetry.
 In Table~\ref{energy}, we list the presently calculated energy levels for some low-lying states of Cs atoms, 
 together with other experimental results from the National Institute of Standards and Technology (NIST) tabulation~\cite{Kramida2015}. 
 Good agreement is obtained.

\begin{table}
	\caption{\label{energy} The presently calculated energy levels~(cm$^{-1}$) for some low-lying states of Cs atoms, as compared with other available experimental results~\cite{Kramida2015}. The binding energies are given relative to the energy of the Cs$^{+}$ core. The experimental data are taken from the National Institute of Science and Technology (NIST) tabulation. The symbol `Diff.' represents the difference of the experimental energies and the present RCICP result.}
	\begin{ruledtabular}
		\begin {tabular}{lcllr}
		States    &  $j$   & RCICP         & Expt.~\cite{Kramida2015} & Diff. \\
		\hline
		$ 6s$     &  1/2  &   $-$31406.5      &  $-$31406.5 &  0.0\\
		$ 6p$    &  1/2  &   $-$20228.2      &  $-$20228.2 &  0.0 \\
	  	&  3/2   &   $-$19674.2                   &  $-$19674.2 &  0.0 \\
		$ 5d$   &  3/2  &   $-$16907.2      &  $-$16907.2 &  0.0 \\
		&  5/2  &   $-$16809.6                   &  $-$16809.6 &  0.0  \\
		$ 7s$  &  1/2  &   $-$12850.2       &  $-$12870.9 & $-$20.7 \\
		$ 7p$  &  1/2  &   $-$9633.9        &  $-$9641.1 & $-$7.2  \\
		&  3/2  &   $-$9452.4                    &  $-$9460.1 & $-$7.7  \\
		$ 6d$  &  3/2  &   $-$8768.9      &  $-$8817.6 & $-$48.7\\
		&  5/2  &   $-$8726.9                  &  $-$8774.8 &$-$47.9  \\
		$ 8s$  &  1/2  &   $-$7078.5      &  $-$7089.3 & $-$10.8\\
		$ 4f$  &  5/2  &   $-$6934.4      &  $-$6934.2 &  0.2  \\
		&  7/2  &   $-$6934.2                  &  $-$6934.4 & $-$0.2 \\
		$ 8p$  &  1/2  &   $-$5694.3     &  $-$5697.6 & $-$3.3 \\
		&  3/2  &   $-$5610.1                  &  $-$5615.0 & $-$4.9 \\
		$ 7d$  &  3/2  &   $-$5328.7     &  $-$5358.6 & $-$29.9  \\
		&  5/2  &   $-$5308.2                  &  $-$5337.7 & $-$29.5 \\
		$ 9s$  &  1/2  &   $-$4488.5      &  $-$4495.8 & $-$7.3 \\
		$ 5f$  &  5/2  &   $-$4435.3      &  $-$4435.2& 0.1  \\
		&  7/2  &   $-$4435.1                &  $-$4435.3 & $-$0.2 \\
		$ 5g$  &  7/2  &   $-$4398.4     &  $-$4398.4 &  0.0  \\
	 	&  9/2  &   $-$4398.4                 &  $-$4398.4 &  0.0   \\
           $ 9p $ & 1/2 & $-$3768.3        &   $-$3769.5    & $-$1.2         \\
           &3/2& $-$3722.6 &$-$3724.8 &$-$2.2 \\
		\end{tabular}
\end{ruledtabular}
\end{table}

\normalsize

\begin{table*}
\caption{\label{matrix1} Comparison of the presently calculated reduced E1 matrix elements (a.u.) and line strength ratios for principal transitions of Cs atoms with other available experimental and theoretical results. The uncertainties are given in parentheses.}
\begin{ruledtabular}
\begin {tabular}{llllll}
 Transitions             & RCICP       & All-order& $\textit{Z}_{\rm final}$~\cite{Safronova2016}   & Theo.         & Expt.\\
                         &             & SDpTsc~\cite{Safronova2016}     &                                        &               &  \\
\hline
$6s_{1/2}\rightarrow 6p_{1/2}$  & 4.5010(77)   &4.5302  &  4.5350(768)   & 4.478~\cite{Safronova1999}, 4.512~\cite{Roberts2014}      & 4.4890(65)~\cite{Rafac1999}, 4.5064(47)~\cite{Derevianko2002}  \\
&&&&&4.508(3)~\cite{Gregoire2016}\\
                         &             &          &           & 4.489~\cite{Kien2013}, 4.510~\cite{Blundell1992}, 4.535~\cite{Safronova1999}             & 4.5010(35)~\cite{Patterson2015}, \textbf{4.5057(16)}~\cite{Damitz2019}  \\
$6s_{1/2}\rightarrow 6p_{3/2}$  & 6.3405(108)   & 6.3734   &  6.3818(789)   & 6.298~\cite{Safronova1999}, 6.347~\cite{Blundell1992}           & 6.3238(73)~\cite{Rafac1999}, 6.3425(66)~\cite{Derevianko2002}  \\              
                         &             &          &           & 6.351~\cite{Roberts2014}, 6.324~\cite{Kien2013}, 6.382~\cite{Safronova1999}                                    & \textbf {6.3349(48)}~\cite{Patterson2015}, 6.345(3)~\cite{Gregoire2016}\\
$6s_{1/2}\rightarrow 7p_{1/2}$	 & 0.2740(55)    & 0.2978   & 0.2983(193)    & 0.2724~\cite{Roberts2014}, 0.2769~\cite{Porsev2010}         &  0.2825(20)~\cite{Shabanova1979}\\
                         &             &          &           & 0.279~\cite{Safronova1999}, 0.279~\cite{Blundell1991}          & 0.2789(16)~\cite{Antypas2013}\\
                         &             &          &           & 0.280~\cite{Blundell1992}, 0.276~\cite{Kien2013}  & 0.2757(20)~\cite{Vasilyev2002}, \textbf {0.27810(45)}~\cite{Damitz2019}\\
$6s_{1/2}\rightarrow 7p_{3/2}$  & 0.5713(114)    & 0.6009   & 0.6013(257)    & 0.5659~\cite{Roberts2014}, 0.586~\cite{Kien2013}                                   & 0.5795(100)~\cite{Shabanova1979}, 0.5856(50)~\cite{Vasilyev2002}\\
                         &             &          &           & 0.576~\cite{Safronova1999},  0.576~\cite{Blundell1992}, 0.575~\cite{Blundell1991}          & 0.5780(7) ~\cite{Antypas2013}, \textbf {0.57417(57)}~\cite{Damitz2019}\\
$6s_{1/2}\rightarrow 8p_{1/2}$	 & 0.0757(27)   & 0.0887   & 0.0916(103)   & 0.081~\cite{Kien2013}, 0.078~\cite{Blundell1992}          &{\textbf{0.072(4)$^a$}~\cite{Kramida2015}}\\
                        &             &          &         &            0.081~\cite{Safronova1999}, 0.092~\cite{Toh2019}                             &  \\
$6s_{1/2}\rightarrow 8p_{3/2}$  & 0.2126(77)    & 0.2282   & 0.2320(144)    & 0.214~\cite{Blundell1992}, 0.232~\cite{Toh2019}            & {\textbf{0.210(8)$^a$}~\cite{Kramida2015}}\\
                       &              &            &       &                0.218~\cite{Safronova1999}, 0.218~\cite{Kien2013}                            &\\
$7s_{1/2}\rightarrow 6p_{1/2}$  & 4.2400(72)    & 4.2313   & 4.2434(121)    &  4.2450~\cite{Porsev2010}     & \textbf {4.249(4)}~\cite{toh2019A} \\
                         &             &          &           &  4.228~\cite{Blundell1991}, 4.236~\cite{Blundell1992}                                   & \\
$7s_{1/2}\rightarrow 6p_{3/2}$  & 6.4746(110)   & 6.4658   &6.4795(187)     & 6.451~\cite{Blundell1991}   & \textbf {6.489(5)}~\cite{toh2019A} \\
                         &             &          &           &  6.470~\cite{Blundell1992}, 6.470~\cite{Kien2013}          &\\
$7s_{1/2}\rightarrow 7p_{1/2}$  &10.324(18)    & 10.2965  &10.3100(401)     &  10.289~\cite{Blundell1992} &\textbf {10.308(15)}~\cite{Bennett1999, Safronova1999} \\
$7s_{1/2}\rightarrow 7p_{3/2}$	 & 14.344(24)   & 14.3028  &14.3231(612)    &  14.293~\cite{Blundell1992} &\textbf {14.320(20)}~\cite{Bennett1999, Safronova1999}\\
$7s_{1/2}\rightarrow 8p_{1/2}$  & 0.9188(55)    & 0.9412   &\textbf{0.9144(268)}     &  0.914~\cite{Toh2019}, 0.935~\cite{Blundell1992}        &\\
$7s_{1/2}\rightarrow 8p_{3/2}$  & 1.6309(98)   & 1.6556   &\textbf{1.6204(352)}     & 1.62~\cite{Toh2019}, 1.647~\cite{Blundell1992}         &\\
$7s_{1/2}\rightarrow 9p_{1/2}$  & 0.3374(67)    & 0.3494   &\textbf{0.3485(101)}     & 0.349~\cite{Toh2019}, 0.375~\cite{Blundell1992}       &\\
$7s_{1/2}\rightarrow 9p_{3/2}$  & 0.6677(134)    & 0.6810   &\textbf{0.6799(142)}     &  0.68~\cite{Toh2019}, 0.725~\cite{Blundell1992}         &\\
$\frac{|\langle 6s_{1/2}||D||6p_{3/2} \rangle|^{2}}{|\langle 6s_{1/2}||D||6p_{1/2} \rangle|^{2}}$& 1.9844(100) &1.979&1.9803(720)&1.9817~\cite{Blundell1992}, 1.9760~\cite{Johnson1996}&1.9845(69)~\cite{Rafac1999}, 1.9809(37)~\cite{Patterson2015}\\
                         &             &          &          &1.9871~\cite{Johnson1987}                     &2.075(8)~\cite{Young1994}\\
$\frac{|\langle 6s_{1/2}||D||7p_{3/2} \rangle|^{2}}{|\langle 6s_{1/2}||D||7p_{1/2} \rangle|^{2}}$&4.3474(2463)& 4.0715&4.063(554)&&  4.5116(876)~\cite{Vasilyev2002}\\
                         &             &          &           &                                 & 4.2950(436)~\cite{Antypas2013}, 4.2626(396)~\cite{Damitz2019}\\
\end{tabular}
\end{ruledtabular}
\label{mat}
\begin{tablenotes}
        \footnotesize
        \item{a:}  These values in NIST~\cite{Kramida2015} are from Ref.~\cite{Morton2000}, which are average values of the experimental results in Ref.~\cite{EXTON1976309, PICHLER1976}.
      \end{tablenotes}
\end{table*}

\subsection{Dipole transition matrix elements}
\begin{table*}
\caption{\label{lifetime} Radiative lifetimes~(ns) and E1 matrix elements $T$ (a.u.) of the 6$p_{1/2}$ and 6$p_{3/2}$ states of Cs atoms. 
The uncertainties are given in parentheses.}          
\begin{ruledtabular}
\begin {tabular}{lllll}
Methods & \multicolumn{2}{c}{ $6p_{1/2}$ }    &	\multicolumn{2}{c}{ $6p_{3/2}$}       \\
\cline{2-3}\cline{4-5}
        & Lifetime   & $T$ &Lifetime & $T$ \\
		\hline
RCICP                                                                                                 & 34.884(119)   &    4.5010(77) &   30.408(104)     &   6.3405(108)   \\
Time-resolved laser \cite{Young1994}                                   & 34.75(7)        &   4.5096(45)                                           &   30.41(10)       & 6.3403(103) \\
Fast-beam laser \cite{Rafac1994}                                             & 34.934(94)   &   4.4978(61)                                           &   30.499(70)     & 6.3311(2) \\
Van der Waals coefficient C6 \cite{Bouloufa2007}           & 34.82(36)     &  4.5051(23)                                            &   30.41(30)       & 6.340(31) \\
Fast-beam laser \cite{Rafac1999}                                             & 35.07(10)     &   4.4890(64)                                          &   30.57(7)          & 6.3236(71) \\
Ultrafast excitation and ionization \cite{Sell2011}            &                       &                                                                &   30.460(38)       & 6.3351(39)\\
Ultrafast excitation and ionization \cite{Patterson2015}    &                      &                                                                 &   30.462(46)       & 6.3349(41) \\
Fast-beam laser \cite{Livingston1992}                                    &                      &                                                                &   30.55(27)         & 6.3258(81)  \\
	\end{tabular}
\end{ruledtabular}
\end{table*}
\normalsize

The E1 transition matrix element can be expressed as
\begin{eqnarray}
	T_{ij}=|\langle \gamma_iJ_i \| \bm D \|\gamma_jJ_i \rangle |,
\end{eqnarray}
\normalsize
where $\gamma$ represents all additional quantum numbers in addition to the total angular momentum $J$. 
$\bm D$ is the dipole transition operator.
In the present calculations, $T_{ij}$ is computed with the use of a modified transition operator~\cite{Mitroy1988, Marinescu1994, Hameed1968,Caves1972,Hafner1978}
\begin{eqnarray}
 \bm{D} =\bm{r}-[1-{\rm exp}(\frac{-r^{6}}{\rho^{6}})]^{1/2}\frac{\alpha^{(1)}_{\rm core}\bm{r}}{r^3},
\label{1}
\end{eqnarray}
where the cutoff parameter $ \rho $=3.6280 a.u. is chosen to ensure that the  presently calculated 
static polarizability equals the measured value 400.8(4) a.u.~\cite{Gregoire2015}.

In Table~\ref{mat}, we list the presently calculated reduced matrix elements for a number of principal E1 transitions of Cs atoms, 
which are compared with the all-order single, double, and triple excitations scaled values (SDpTsc)~\cite{Safronova2016} 
as well as other theoretical and experimental results. 
The column labeled by $\textit{Z}_{\rm{final}}$ shows the 
final results of Ref.~\cite{Safronova2016} calculated with inclusion of higher-order correlation. 
For the $6s_{1/2} \rightarrow 6p_{1/2,3/2}$ transitions, the present results differ from the results of $\textit{Z}_{\rm final}$ by about 0.8\%. 
Overall, the present RCICP results are in excellent agreement with 
the experimental results from Young \emph{et al.}~\cite{Young1994} and Patterson \emph{et al.} ~\cite{Patterson2015}. 
As for the $6s_{1/2} \rightarrow 7p_{1/2,3/2}$ transitions, the present results are in good agreement with 
the experimental results in Refs.~\cite{Vasilyev2002, Antypas2013, Shabanova1979, Damitz2019}. 
Furthermore, the present results for the $7s_{1/2} \rightarrow np_j (n=6, 7, 8)$  
transitions agree very well with the results in Refs.~\cite{Safronova2016, Arora2007, Blundell1992, Dzuba1989, Toh2019}.  
Uncertainties of the present results, as well as those of \textit{Z}$_{\rm final}$, 
are given in Table~\ref{mat} in parentheses after the values in units of the last digit of the value. 
The method of estimation of the uncertainties of our calculations is explained in Section IV.

Besides the E1 reduced matrix elements, the ratio of the line strengths~(i.e., squares of the E1 matrix elements) 
of the $6s_{1/2} \rightarrow 6p_{1/2}$ and $6s_{1/2} \rightarrow 6p_{3/2}$ transitions are also given in Table~\ref{mat}. 
Although this ratio should be exactly 2.0 in the nonrelativistic limit, both the present and other available results indicate 
that it is slightly smaller than 2.0. The reason for such a deviation from 2.0 is the spin-orbital splitting of Cs atoms. 
Nevertheless, the present RCICP ratio 1.9844(100) agrees excellently with the experimental value 1.9845(69)~\cite{Rafac1999}. 
In contrast, the corresponding ratio for the $6s_{1/2} \rightarrow 7p_{1/2,3/2}$ transitions differs substantially from 2.0, 
but the present RCICP result is still in good agreement with the experimental 
results~\cite{Shabanova1979, Vasilyev2002, Antypas2013, Damitz2019} 
and  differs from $\textit{Z}_{\rm{final}}$ of Ref.~\cite{Safronova2016} only by about 6\%.

As there is only one spontaneous decay channel $6p_j \rightarrow 6s_{1/2}$ for each of the $6p_j$ states, 
experimentally measured lifetimes of the states can be used to determine the transition matrix elements. 
Table~\ref{lifetime} lists the presently calculated lifetimes of the $6p_{1/2}$ and $6p_{3/2}$ states together with 
other experimental results~\cite{Rafac1999, Young1994, Rafac1994, Bouloufa2007, Sell2011, Patterson2015, Livingston1992}. 
The present RCICP lifetime of the $6p_{1/2}$ state is smaller than the experimental result~\cite{Rafac1999} by about 0.6\%. 
Nevertheless, it is still in good agreement with other experimental results~\cite{Bouloufa2007, Rafac1994, Young1994}. 
For the $6p_{3/2}$ state, the present lifetime agrees well with the experimental 
results from Refs.~\cite{Young1994, Rafac1994, Bouloufa2007, Sell2011, Patterson2015}, 
while it is smaller than the experimental results of Tanner \emph{et al.}~\cite{Livingston1992} and of Rafac \emph{et al.}~\cite{Rafac1999} just by about 0.3\%.

\subsection{Polarizabilities of the ground state}
The scalar dynamic polarizability of the ground state is written as~\cite{mitroy2010}
\begin{eqnarray}
\alpha^S (\omega) =\sum_{n} \frac{f_{0 \rightarrow n}}{(\Delta E_{0\rightarrow n}^2- \omega ^2)},
\label{3}
\end{eqnarray}
where $\Delta E_{0\rightarrow n}=E_{n}-E_{0}$ is the transition energy, 
the summation over $n$ includes all allowable fine-structure transitions, $\omega$ is the frequency of the external field.  
$f_{0 \rightarrow n}$ denotes the oscillator strength of a dipole transition, which is defined as
\begin{eqnarray}
	f_{0 \rightarrow n}=\frac{2|\langle \gamma_nJ_n \| \bm D \|\gamma_0J_0 \rangle |^2\Delta E_{0\rightarrow n}}{3\times(2J_0+1)}.
\label{7}
\end{eqnarray}
If $\omega=0$, the dynamic polarizability in Eq. (\ref{3}) is reduced to the static polarizability.  
By using the experimental energy levels and E1 matrix elements obtained above, the static dipole polarizability can be readily calculated. 
In Table~\ref{vb}, we list the presently calculated static dipole polarizabilities of the ground state of Cs atoms together with other available results. 
The present RCICP results agree very well with the theoretical results 
of R-$ab$-$initio$~\cite{Derevianko1999}, RCC-SD~\cite{Derevianko2002, Singh2016}, 
R-SD~\cite{Derevianko1999, Safronova1999} and R-all-order~\cite{Iskrenova2007} methods.

\begin{table}
	\caption{\label{polarizability} Dipole static polarizabilities (a.u.) of the ground state of Cs atoms.}
	\begin{ruledtabular}
		\begin{tabular}{ll}
			&  \multicolumn{1}{c}{$ \alpha^S~(\rm a.u.)$} \\
			\hline	
			RCICP                                                                  &  400.80(97)               \\
			DK+CCSD(T)$^a$~\cite{Lim2005}             &  396.02                  \\
			RCCSD(T)$^b$~\cite{Borschevsky2013}   &  399.0                   \\
			R-SD$^c$~\cite{Safronova1999}                  &  399.9                   \\
			R-SD$^c$~\cite{Derevianko1999}              &  401.5                   \\
			R all-order$^d$~\cite{Iskrenova2007}      &  398.4(7)                \\
			RCC+SD$^e$~\cite{Singh2016}                  &  399.5(8)               \\
			RCC+SD$^e$~\cite{Derevianko2002}       &  400.49(81)         \\
			CCSD(T)$^f$~\cite{Lim1999}                     &  430                  \\
	      	R $ab$~$initio$ $^g$~\cite{Derevianko1999}   &  399.9(1.9)      \\
			Expt.~\cite{Amini2003}                                 &  401.0(6)         \\
		 	Expt.~\cite{Gregoire2015}                            &  400.8(4)         \\
		\end{tabular}
		\label{vb}
	\end{ruledtabular}
\begin{tablenotes}
        \footnotesize
        \item{a:}  Douglas-Kroll approximation combined with the coupled cluster method with single, double, and perturbative triple
excitations  (DK+CCSD(T)).
        \item{b:}  Relativistic single reference coupled cluster approach with single, double, and perturbative triple excitations (RCCSD(T)).
        \item{c:}  Relativistic single-double R-SD.
        \item{d:}  Relativistic all-order method (R all-order).
        \item{e:}  Relativistic coupled-cluster singles and doubles method RCC+SD.
        \item{f:}  Single, double, and perturbative triple excitations (CCSD(T)).
        \item{g:}  Relativistic $ab$ $initio$ methods (R $ab$ $initio$).
      \end{tablenotes}
\end{table}

\begin{table*}
\caption{Breakdown of the contributions to the static and dynamic polarizabilities 
(in a.u.) at three longest tune-out wavelengths $\lambda_{\rm zero}$ of  the ground state of Cs. 
The remainder term ``$Remainder$'' comes from all of the valence transitions except those listed specifically in the table.   
$\lambda_{\rm zero}^{recom}$ represents the results that are obtained with a replacement of 
the RCICP $6s_{1/2}\rightarrow 6p_{1/2,3/2}$ and $6s_{1/2}\rightarrow 7p_{1/2,3/2}$ matrix elements by 
the experimental results~\cite{Damitz2019,Patterson2015}.  
The uncertainties are given in parentheses.}
\label{tuneout}
\begin{ruledtabular}
\begin {tabular}{lllll}
$\lambda_{\rm zero}$~(nm)        &    $\infty$      &  880.2510(475)     &   460.1897(543)      &   457.2428(917)     \\
$\lambda_{\rm zero}^{recom}$~(nm)  &     &  880.2144(158) &   460.2154(63) &   457.2504(171)\\
Ref.~\cite{Arora2011}              &                  &   880.25(4)         &  460.22(2)           &   457.31(3)            \\
Ref.~\cite{LeBlanc2007}          &                  &   880.29              &                               &               \\
\hline
$6s_{1/2}\rightarrow 6p_{1/2}$  &   132.587(451)      & $-$4035.953    &   $-$47.710         &   $-$46.886      \\
$6s_{1/2}\rightarrow 6p_{3/2}$  &   250.683(852)      & 4017.709          &   $-$103.140       &  $-$101.291   \\
$6s_{1/2}\rightarrow 7p_{1/2}$  &   0.252(10)            & 0.347                 &   78.101               &  $-$26.115      \\
$6s_{1/2}\rightarrow 7p_{3/2}$  &   1.088(44)            &  1.486                &  55.490              &  157.005         \\
$6s_{1/2}\rightarrow 8p_{1/2}$  &  0.016(1)               &  0.020                &  0.057                  & 0.059             \\
$6s_{1/2}\rightarrow 8p_{3/2}$  & 0.128(9)                &  0.159                &    0.441                & 0.456             \\
Remainder                                       &  0.247(25)             &  0.269               &  0.386                   & 0.390             \\
$\alpha_{\rm core}$       &   15.8(1)~\cite{Lim2002}  & 15.964              &    16.376              & 16.383           \\
Total                                                 &   400.80(97)           &   0                     &    0                         &  0                    \\
\end{tabular}
\end{ruledtabular}
\end{table*}
\normalsize

\subsection{Tune-out wavelengths}

FIG.~1 shows the dynamic polarizabilities of the ground state of Cs atoms. 
Three visible tune-out wavelengths are found, as indicated by arrows. 
For the present atomic system, the tune-out wavelengths appear in two cases. 
In the first case, the tune-out wavelengths lie between the $6s \rightarrow np_{j} (n=6,7)$ fine-structure components. 
For example, the tune-out wavelength 880.2510~nm lies between the $6s_{1/2} \rightarrow 6p_{1/2,3/2}$ resonances, 
while the one 457.2428~nm lies between the $6s_{1/2} \rightarrow 7p_{1/2,3/2}$ resonances. 
In the second case, the tune-out wavelength lies between 
the $6s_{1/2} \rightarrow np_{3/2}$ and $6s \rightarrow (n+1)p_{1/2} (n=6, 7\dots)$ transitions. 
For example, the tune-out wavelength 460.1897~nm lies between 
the $6s_{1/2} \rightarrow 6p_{3/2}$ and $ 6s_{1/2} \rightarrow 7p_{1/2}$ transitions. 
In comparison, the present RCICP tune-out wavelengths are found to be very close to 
the theoretical results of Arora \emph{et al.}~\cite{Arora2011}, as seen from Table~\ref{tuneout}.

Besides the tune-out wavelengths, Table~\ref{tuneout} also lists the breakdown of the contributions to the
dynamic polarizabilities at the tune-out wavelengths. 
It is found that the 880-nm tune-out wavelength is mainly caused by a cancellation of 
the contributions from the $6s_{1/2} \rightarrow 6p_{1/2}$ and $6s_{1/2} \rightarrow 6p_{3/2}$ transitions; 
the contributions from other transitions are negligibly small. At the tune-out wavelengths near 460 nm and 457 nm, 
the main contributions to the dynamic polarizabilities are from 
the transitions from the $6p_j$ and $7p_j$ states. 
These tune-out wavelengths can be effectively determined by means of a relative 
magnitude of the $6s_{1/2} \rightarrow 6p_{1/2,3/2}$ and $6s_{1/2} \rightarrow 7p_{1/2,3/2}$ matrix elements or oscillator strengths.

Since more accurate experimental results 6.3349(48) a.u., 4.5057(16) a.u., 0.27810(45) a.u., and 0.57417(57) a.u. of the $6s_{1/2} \rightarrow 6p_{1/2,3/2}$ and $6s_{1/2} \rightarrow 7p_{1/2,3/2}$ transition matrix elements can be obtained from Ref.~\cite{Patterson2015, Damitz2019}, we recalculate the tune-out wavelengths by replacing the corresponding RCICP matrix elements with these experimental values, which are labeled as $\lambda_{\rm zero}^{recom} $ in Table~\ref{tuneout}.  
Evidently,  the uncertainties of  $\lambda_{\rm zero}^{recom}$ are much smaller than those of
the RCICP tune-out wavelengths. 
Therefore, in the calculations of polarizabilities and tune-out wavelengths for the hyperfine components of the ground level in Sec.~III,  the 
RCICP matrix elements of the $6s_{1/2} \rightarrow 6p_{1/2,3/2}$ and $6s_{1/2} \rightarrow 7p_{1/2,3/2}$ transitions are replaced by the experimental values.  
Also, the RCICP calculation results are listed in the Supplemental Materials~\cite{Suppl}.

\begin{figure}[tbh]
\vspace{-0.4em}
\label{dynamic}
\setlength{\abovecaptionskip}{0pt}
\setlength{\belowcaptionskip}{0pt}
\centering{
\includegraphics[width=10cm,height=8cm]{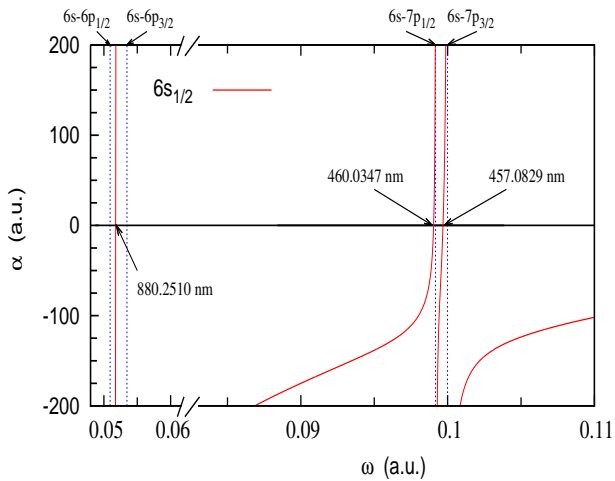}}
\vspace{0.5em}
\caption{Dynamic polarizabilities of the ground state of Cs atoms. Three visible tune-out
wavelengths are found, which are marked with arrows. Approximate positions of the
$6s_{1/2}\to np_j$ resonances are indicated by vertical dotted lines.}
\end{figure}

\section{Calculations for the hyperfine components of the ground level of $^{133}$Cs}

\subsection{Energies and reduced matrix elements}
In order to further consider the effect of the hyperfine interaction on the polarizabilities, 
the theory of energies and reduced matrix elements should be reformulated. The energy of a hyperfine state is given by
\begin{eqnarray}
	E^{\rm hf}=E_J+W_F,
\end{eqnarray}
where $E_J$ is the energy of an unperturbed fine-structure level. 
$W_F$ represents the hyperfine interaction energy, which can be expressed as
\begin{eqnarray}
	W_F=\frac{1}{2}AR+B\frac{\frac{3}{2}R(R+1)-2I(I+1)J(J+1)}{2I(2I-1)2J(2J-1)},
\label{WF}
\end{eqnarray}
where \textit{A} and \textit{B} denote the hyperfine structure constants, 
$I$ is the nuclear spin ($I$ = 7/2 for $^{133}$Cs), and $J$ is total angular momentum of fine-structure energy levels. $R$ is given by
\begin{eqnarray}
	R=F(F+1)-I(I+1)-J(J+1)
\end{eqnarray}
with total angular momentum $F$ of hyperfine-structure energy levels.

In Table~\ref{ty}, we list the hyperfine structure constants {\textit{A} and \textit{B}, 
the hyperfine interaction energies, and the hyperfine energy levels. 

\begin{table*}
	\caption{\label{tab14} The hyperfine structure constants \textit{A} and \textit{B}, hyperfine interaction energies, and hyperfine energies of some low-lying states of $^{133}$Cs atoms. The notation a[b] represents a$\times10^b$.}
	\begin{ruledtabular}

		\begin {tabular}{ccllccc}
		States    &  $J$   & \textit{A} (MHz)     &  \textit{B} (MHz)              &   \textit{F}              & $W_F$ (a.u.)             & $E^{\rm {hf}}$ (a.u) \\
		\hline
		$ 6s$   &  1/2  &   2298.157943~\cite{Williams2018,Carr2016}  &                    &  3               &  $-$7.858[$-$7]   & $-$0.14309918241 \\
		        &       &                 &                    &  4               &  $+$6.112[$-$7]   & $-$0.14309778529  \\
		$ 6p$   &  1/2  &   291.9309(12)~\cite{Gerginov2006}     &                    &  3               &  $-$9.982[$-$8]   & $-$0.09216655791   \\
                 	 &       &       &                              &  4               &  $+$7.764[$-8$]   & $-$0.09216638044  \\
	                  & 3/2   &   50.28825(23)~\cite{Johnson2004}      &   $-$0.4940(17)~\cite{Johnson2004,Carr2016}       &   2              &  $-$5.163[$-$8]   & $-$0.08964212213  \\
                       &       &                 &                  & 3                &  $-$2.865[$-8$]   & $-$0.08964209915  \\
	        	      &       &                 &                  & 4                &  $+$1.946[$-9$]   & $-$0.08964206855  \\
	        	      &       &                 &                  & 5                &  $+$4.011[$-8$]   & $-$0.08964203039 \\
	    $ 7p$   &  1/2  &   94.35(4)~\cite{Feiertag1972Core}        &                     & 3                   &  $-$3.226[$-$8]   & $-$0.04389478454 \\
	  	           &  &     &&                       4                               &  $+$2.509[$-$8]   & $-$0.04389472718 \\
	           	     &  3/2  &   16.609(5)~\cite{Belin1976}    & $-$0.15(3)~\cite{S2005Level} &    2                            &  $-$1.705[$-$8]   & $-$0.04306790886 \\      	
	                &        &&&                                 3                    &  $-$9.460[$-$8]   & $-$0.04306790127  \\
	                &&&                                       &  4                    &  $+$0.642[$-$9]   & $-$0.04306789117  \\
	                &&                &                       &   5                    &  $+$1.324[$-$8]   & $-$0.04306787856    \\	
	\end{tabular}
	\label{ty}
\end{ruledtabular}
\end{table*}
\normalsize

By means of the Wigner-Eckart theorem, the transition matrix elements between two
 hyperfine states $| \gamma_i J_i I F_i\rangle $ and $ |\gamma_g J_g I F_g\rangle$ can be written as
\begin{eqnarray}
	&T^{\rm hf}_{ig}=\langle \gamma_i J_iIF_i \| \bm D\| \gamma_g J_gIF_g\rangle  \nonumber \\
      &= (-1)^{I+J_g+F_i+1}
	 \times  \hat{F}_i \hat{F}_g
	\left\{
	\begin{array}{ccc}
		I & J_i & F_i \\
		1 & F_g & J_g \\
	\end{array}
	\right \}
	T_{ig}.
\label{2}
\end{eqnarray}
In this expression, $\hat{F}=\sqrt{2F+1}$. 

\begin{table}
\caption{\label{tab16} Partial derivatives (in a.u.) of the matrix elements of the $6s_{1/2}-6p_{1/2,3/2}$ 
transitions with respect to the initial- and final-state binding energies.
The uncertainties are given in parentheses.}

\begin{ruledtabular}
\begin {tabular}{ccc}
Transitions                              &     $\frac{\partial T}{\partial E_{6s}}$  &  $\frac{\partial T}{\partial E_{j}}$ \\\hline
6$s_{1/2}\rightarrow 6p_{1/2}$         &  34.82(42)      &    $-$0.98(4)                     \\
6$s_{1/2}\rightarrow 6p_{3/2}$         & 51.76(62)       &    $-$7.50(30)                  \\
\end{tabular}
\end{ruledtabular}
\end{table}

\subsection{Dipole polarizabilities}
\normalsize
The calculation of dipole scalar polarizabilities for the hyperfine states is rather similar to Eq.~(\ref{3}), 
except that $f_ {0\rightarrow n}$ should be replaced by the hyperfine oscillator strengths $f^{\rm hf}_{g\rightarrow i}$, 
and $\Delta E_{0\rightarrow n}$ is replaced by the hyperfine transition energy $\Delta E_{g\rightarrow i}$. 
Likewise, the calculation of hyperfine absorption oscillator strength $f^{\rm hf}_ {g\rightarrow i}$ is similar to Eq.~(\ref{7}) 
except that $J_0$ should be replaced by $F$ and the matrix element is replaced by Eq.~(\ref{2}). 
Because the total angular momentum of the hyperfine components of the ground level is greater than one, 
the corresponding dipole polarizability has a tensor component, which can be written as
\begin{eqnarray}
&&\alpha^{\rm T}(\omega)=6\left( \frac{5F_g(2F_g-1)(2F_g+1)}{6(F_g+1)(2F_g+3)}
\right)^{1/2} \nonumber \\
& \times & \sum_{i}  (-1)^{F_g+F_i}
\left\{
\begin{array}{ccc}
F_g & 1   & F_i \\
1 & F_g & 2  \\
\end{array}
\right\}
\frac{ f^{\rm hf}_{g \rightarrow i} } {\Delta E_{g\rightarrow i}^2 - \omega^2}.
\end{eqnarray}
Here, $\Delta E_{g\rightarrow i}=E^{\rm hf}_i-E^{\rm hf}_g$. The total polarizability of a hyperfine state is written as
\begin{equation}
	\alpha_{M_{F_g}}(\omega) = \alpha^S(\omega) + \alpha^{T}(\omega) \frac{3M_g^2-F_g(F_g+1)}{F_g(2F_g-1)},
	\label{polar3}
\end{equation}
in which $M_{F_g}$ is the magnetic component of the quantum number $F_g$.

In the calculations of polarizabilities, 
small corrections to the matrix elements of the $6s_{1/2} \rightarrow 6p_{1/2,3/2}$ transitions 
are made, which are treated as parametric functions of the binding energies of the lower and upper levels and can be written as
\begin{eqnarray}
T^{c}_{6s_{1/2},{6p_{j}}} \approx& T_{6s_{1/2},{6p_{j}}} 
+ \frac{\partial T_{{6s_{1/2}},6p_{j}}}{\partial E_{6s_{1/2}}}W_{F_{6s_{1/2}}}& \nonumber \\
&+ \frac{\partial T_{{6s_{1/2}},{6p_{j}}}}{\partial E_{6p_{j}}}W_{F_{6p_{j}}},&
\end{eqnarray}
where $T_{6s_{1/2},6p_j}$ denotes the $6s_{1/2} \rightarrow 6p_{1/2,3/2}$ transition matrix elements, and $W_F$ are defined by Eq.~\ref{WF}.
The partial derivatives denote the ratio of the changes in the reduced matrix elements and binding energies.  
In the present calculations,  we repeat the calculations of these derivatives many times, each time using 
a slightly different polarization potential (with the change of binding energy in the range $10^{-9}$-$10^{-5}$ a.u.).  
Table~\ref{tab16} lists the average values of these partial derivatives and the corresponding uncertainties that are 
determined as root-mean-squares (rms) of the differences from the average values.
Then, the dipole polarizabilities for the hyperfine states are further determined by using these partial derivatives.

Table~\ref{tn} lists the dipole scalar and tensor polarizabilities of the hyperfine components of the ground level of $^{133}$Cs. 
The uncertainties are derived from the uncertainties of the partial derivatives in Table~\ref{tab16}.
The tensor polarizabilities of the hyperfine states are found to be $10^{-4}$ in magnitude. 
As far as we know, there are no other theoretical or experimental results for direct comparison. 
Nevertheless, we can compare the present hyperfine Stark shift of the ground states 
with other available results~\cite{Feichtner1965, Lee1975, Palchikov2003, Micalizio2004, 
Angstmann2006, Haun1957, Mowat1972, Bauch1997, Simon1998, Levi2004, Godone2005}. 
It is often reported in experiment as a hyperfine Stark shift coefficient $k$~(in Hz/(V/m)$^2$),  
i.e., $k =-0.5\Delta \alpha$ with the difference $\Delta \alpha$ in the scalar polarizabilities of 
the hyperfine components  $F$=4 and $F$=3 of the ground level. 
Table~\ref{star} lists the hyperfine Stark shift coefficient $k$ and $\Delta \alpha$.
As can be seen from this table, the present hyperfine Stark shift agrees very well with 
other experimental~\cite{Haun1957, Mowat1972, Simon1998, Levi2004, Godone2005, Bauch1997} 
and theoretical results~\cite{Feichtner1965, Palchikov2003, Micalizio2004, Angstmann2006, Beloy2006, Lee1975}.

The correction of Eq. (14) to the $6s_{1/2} \rightarrow 6p_{j}$ 
transition matrix elements is very significant for the result of the hyperfine Stark shift.  
The $\Delta \alpha$ without considering the correction of the transition matrix elements 
is just $-$0.01013 a.u. That is, the correction of the $6s_{1/2} \rightarrow 6p_{1/2,3/2}$ transition matrix elements 
makes the hyperfine Stark shift increase by about 85\%.

\begin{table}
\caption{\label{tn}Dipole scalar $\alpha^S$ and tensor $\alpha^T$ polarizabilities of the hyperfine components of the ground level of $^{133}$Cs. 
The uncertainties are given in parentheses. a[b] represents a$\times10^b$.
}
\begin{ruledtabular}
\begin {tabular}{lccl}
   States     & $F$ &  $\alpha^S$  (a.u)& $\alpha^T$ (a.u.)     \\\hline
 6s$_{1/2}$   & 3 &400.66237(3)   & 1.2263(24)[$-$4]          \\
 6s$_{1/2}$   & 4 &400.68105(3)  & $-$2.2919(45)[$-$4]     \\ 	
\end{tabular}
\end{ruledtabular}
\end{table}

\begin{table}
\caption{\label{star}Hyperfine Stark shift coefficients $k$ (in Hz/(V/m)$^2$) and the differences $\Delta\alpha$ (in a.u.) in the
scalar polarizabilities of the hyperfine components of the ground level $F$=4 and $F$=3. The
unit Hz/(V/m)$^2$ can be converted into a.u. by multiplying $0.4018778 \times 10^8$.
The uncertainties are given in parentheses. }
\begin{ruledtabular}
  \begin {tabular}{lll}
		Methods                  & $\Delta\alpha     $         &$ k~(\times 10^{-10}$ )  \\\hline
     RCICP                                                                   &     0.01868(4)                   &   $-$2.324(5)               \\
     MBPT~\cite{Feichtner1965}                            &     0.0153(24)                 &   $-$1.9(3)   \\
     MBPT~\cite{Lee1975}                                       &     0.017925                    &  $-$2.2302  \\
     MCDF~\cite{Palchikov2003} 	                      &      0.018325                     &   $-$2.28      \\
     $ab~initio$~\cite{Micalizio2004}                   &      0.01583(72)                &    $-$1.97(9) \\
     $ab~initio$~\cite{Angstmann2006} 	          &       0.01817(16)                &   $-$2.26(2)   \\
     Relativistic many-body~ \cite{Beloy2006}   &      0.018253(65)             &   $-$2.271(8)  \\
        Perturbation theory~\cite{2006Stark}         &     0.01656(8)                  & $-$2.06(1)\\
     Expt.~\cite{Haun1957}                                    &       0.01841(56)                &    $-$2.29(7)  \\
     Expt.~\cite{Mowat1972}                                 &       0.01809(40)                &   $-$2.25(5)\\
     Expt.~\cite{Bauch1997}                                   &       0.01744(209)              &  $-$2.17(26) \\
     Expt.~\cite{Simon1998}                                  &       0.018253(32)               &   $-$2.271(4) \\
     Expt.~\cite{Levi2004}                                     &        0.01519 (96)                 &    $-$1.89(12) \\
     Expt.~\cite{Godone2005}                              &        0.01632(32)                  &     $-$2.03(4) \\
\end{tabular}		
\end{ruledtabular}
\end{table}

\begin{table*}
\caption{\label{133cs} Tune-out wavelengths $\lambda_{\rm zero}$ (nm) of the $6s_{1/2}, 
F$=3, 4 states of $^{133}$Cs atoms. $\Delta \lambda$ denotes the shifts of the primary 
tune-out wavelengths relative to the tune-out wavelengths of the $6s_{1/2}$ state. 
$\Delta E$ denotes the corresponding energy shifts.
}
\begin{ruledtabular}
\begin {tabular}{llllll}
		\multicolumn{3}{c}{$F$=3}  &  \multicolumn{3}{c}{$F$=4}     \\
		\cline{1-3} \cline{4-6}
		$\lambda_{\rm zero}$ & $ \Delta \lambda~(10^{-2} $ nm) &  $\Delta E~(10^{-7}$ a.u.)  & $\lambda_{\rm zero}$ &$ \Delta \lambda~(10^{-2}$ nm)  & $ \Delta E~(10^{-7}$ a.u.)   \\\hline
		894.58011017(6)       &        -{}-      &        -{}-      & 894.60361080(6)   &   -{}-        &    -{}-       \\
         	880.2008518(18)      &    $-$1.35482(18)    &       7.9965(10)      & 880.2250251(18)    &  1.06251(18)       &  $-$6.2194(10)        \\
		852.33538210(7)       &      -{}-        &         -{}-     & 852.35737909(7)    &   -{}-        &    -{}-       \\
		852.33485022(7)       &     -{}-         &         -{}-     & 852.35675635(7)   & -{}-          &    -{}-       \\
		460.2117478(5)      &    $-$0.36522(5)     &      7.9440(9)      & 460.2182931(5)   & 0.28931(5)     &   $-$6.1364(9)         \\
		459.44234414(4)       &     -{}-         &        -{}-      & 459.44872828(4)    &  -{}-         &    -{}-       \\
		457.2468549(4)      &   $-$0.35451(4)&       7.7871(8)     & 457.2531880(4)    &  0.27880(4)      &    $-$6.0143(8)       \\
		455.65213817(3)       &      -{}-        &        -{}-      & 455.65847816(3)    & -{}-          &    -{}-       \\
		455.65208799(3)       &     -{}-         &       -{}-       & 455.65841938(3)    &  -{}-         &    -{}-         \\    	
\end{tabular}
\end{ruledtabular}
\end{table*}
\begin{table}
\caption{\label{lli} Tune-out wavelengths $\lambda_{\rm zero}$~(nm) of the magnetic sublevels of the $6s_{1/2}, F$=3, 4 states of $^{133}$Cs atoms. $\Delta \lambda$~(nm) is the difference of the tune-out wavelengths with and without considering the tensor contributions.} 
\begin {tabular}{llll} \hline
\hline
    $\lambda_{\rm zero}$ & $ \Delta \lambda $ & $\lambda_{\rm zero}$&  $ \Delta \lambda $   \\
		\cline{1-4}
\multicolumn{4}{c}{$F=3$}       \\
\multicolumn{2}{c}{ $M_{\rm F}=\pm 3$} &\multicolumn{2}{c}{ $M_{\rm F}=\pm 2$}   \\
880.2009172(18)  &	6.54$\times 10^{-5}$	     &   880.2008518(18) & $\textless 10^{-7}$   \\
460.2117558(5)   &	 8.00$\times 10^{-6}$     &  460.2117478(4)       & $\textless 10^{-7}$ \\  
457.2468553(4)   & 0.40$\times  10^{-6}$    & 457.2468548(4)               &  $\textless 10^{-7}$   \\
	\cline{1-4}
\multicolumn{2}{c}{ $M_{\rm F}=\pm 1$} &\multicolumn{2}{c}{ $M_{\rm F}=\pm 0$}   \\
880.2008126(17)   & $-$3.92$\times 10^{-5}$  & 880.2007995(17) &$-$5.23$\times 10^{-5}$\\
460.2117432(2)   & $-$4.62$\times 10^{-6}$  & 460.2117415(3)    &	$-$6.30$\times 10^{-6}$  \\
457.2468547(2)  &  $-$0.20$\times  10^{-6}$   & 457.2468546(3)  &$-$0.30$\times  10^{-6}$   \\
\cline{1-4}
\multicolumn{4}{c}{$F=4$}       \\
\multicolumn{2}{c}{ $M_{\rm F}=\pm 4$} &\multicolumn{2}{c}{ $M_{\rm F}=\pm 3$}   \\
880.2249277(18) &  $-$9.74$\times 10^{-5}$    &  880.2250007(18)  &  $-$2.44$\times 10^{-5}$  \\
460.2182814(5)  & $-$1.17$\times 10^{-5}$     &  460.2182902(4)    & 3.00$\times 10^{-6}$ \\
457.2531875(4)  &   $-$0.50$\times  10^{-6}$            &  457.2531878(4)    &  $-$0.20$\times  10^{-6}$    \\
\cline{1-4}
\multicolumn{2}{c}{ $M_{\rm F}=\pm 2$} &\multicolumn{2}{c}{ $M_{\rm F}=\pm 1$}   \\
   880.2250529(17) & 2.78$\times 10^{-5}$         & 880.2250842(17) & 6.00$\times 10^{-5}$ \\
  460.2182964(2)  & $-$3.30$\times 10^{-6}$     & 460.2183002(3)  & 7.10$\times 10^{-6}$ \\
  457.2531882(2)  &  0.20$\times  10^{-6}$              & 457.2531883(3)   & 0.20$\times  10^{-6}$   \\
  \cline{1-4}
  \multicolumn{2}{c}{ $M_{\rm F}=\pm 0$} &   \\
  880.2250946(17) &  6.95$\times 10^{-5}$&& \\
  460.2183015(2) &8.40$\times 10^{-6}$&&\\
   457.2531884(2)& 0.40$\times  10^{-6}$    &&\\
   \hline \hline
\end{tabular}
\end{table}
\normalsize

\subsection{Tune-out wavelengths}

By means of the hyperfine energy levels and matrix elements obtained above, 
the dynamic polarizabilities and tune-out wavelengths of the hyperfine components of the ground level can be further determined. 
Table~\ref{133cs} lists the tune-out wavelengths of the hyperfine components of the ground level of $^{133}$Cs, 
in which the tensor contribution is not considered. Two new features of the tune-out wavelengths are found for the hyperfine splitting. 
One of them is that the hyperfine splitting of the ground state results in two sets of tune-out wavelengths; one is for the 6$s_{1/2}, F$ = 3 state, 
and the other is for the 6$s_{1/2}, F = 4$ state. 
Another feature is that the hyperfine splitting of the $np_{j}$ states results in additional tune-out wavelengths. 
The tune-out wavelengths near 894 nm and 852 nm 
are located between the hyperfine components of the transitions from the
$6p_{1/2}$ and $6p_{3/2}$  levels to the ground level, respectively, 
while the ones near 459 nm and 455 nm  
are located between the hyperfine components of the transitions 
from the 7$p_{1/2}$} and 7$p_{3/2}$ levels to the ground level, respectively. 
These tune-out wavelengths would be difficult to be measured in experiments 
due to very small energy splittings of the hyperfine states. 
The tune-out wavelength near 880 nm is the first primary tune-out wavelength, 
which lies between the excitation thresholds of the $6p_{1/2}$ and $6p_{3/2}$ states. 
The primary tune-out wavelength 880.200852~nm of the 
6$s_{1/2}, F = 3$ state is shorter than the 880.2144~nm tune-out wavelength 
of the 6$s_{1/2}$ state by about 0.013548~nm. 
The tune-out wavelength 880.225025~nm of the 6$s_{1/2}, F = 4$ state is longer 
than the 880.2144~nm tune-out wavelength of the 6$s_{1/2}$ state by about 0.010625~nm. 
As can be seen from Table~\ref{133cs}, the energy shifts of the first 
primary tune-out wavelengths for the $6s_{1/2}, F = 3, 4$ states are 
7.997 $\times 10^{-7}$ a.u. and $-$6.219 $\times 10^{-7}$ a.u., respectively. 
These shifts correspond very closely to the hyperfine interaction energies of the $6s_{1/2}$ state. 
The tune-out wavelengths near 460 nm and 457 nm are also primary tune-out wavelengths, 
which lie between the excitation thresholds of the 6$p_{3/2}$ and 7$p_{1/2}$ states and 
the thresholds of the 7$p_{1/2}$ and 7$p_{3/2}$ states, respectively. The energy shifts of 
the tune-out wavelengths are also very close to the hyperfine interaction energy of the ground state.

While the calculations of the hyperfine Stark shift are critically reliant on the use of 
the energy-adjusted reduced dipole matrix elements, this is not true for the tune-out wavelengths. 
For example, the difference of the tune-out wavelengths near 880 nm with and without considering 
the correction to the 6$s_{1/2} \rightarrow 6p_{1/2,3/2}$ transition matrix elements are 
only $-$8.5$\times 10^{-5}$ nm and $-$6$\times 10^{-6}$ nm for the $F = 3$ and $F = 4$ states, respectively. 
The influence of the correction to the transition matrix elements on the tune-out wavelengths can be ignored.

If tensor polarizabilities are considered, the tune-out wavelengths also depend on the magnetic sublevels. 
Table~\ref{lli} lists the primary tune-out wavelengths for the magnetic sublevels of the $6s_{1/2}$ $F = 3$ and $F = 4$ states of $^{133}$Cs.
The differences of the tune-out wavelengths with and without considering the tensor contributions are also listed in Table~\ref{lli}. 
It is found that the differences in these tune-out wavelengths for any of the magnetic sublevels are $10^{-5}\sim10^{-6}$~nm. 
If the tune-out wavelengths could be measured at a level of fm, the contribution of the tensor polarizabilities to 
the tune-out wavelengths could be extracted from experiments.

\begin{figure}[tbh]
\vspace{-0.4em}
\label{dynamic}
\setlength{\abovecaptionskip}{0pt}
\setlength{\belowcaptionskip}{0pt}
\centering{
\includegraphics[width=10cm,height=8cm]{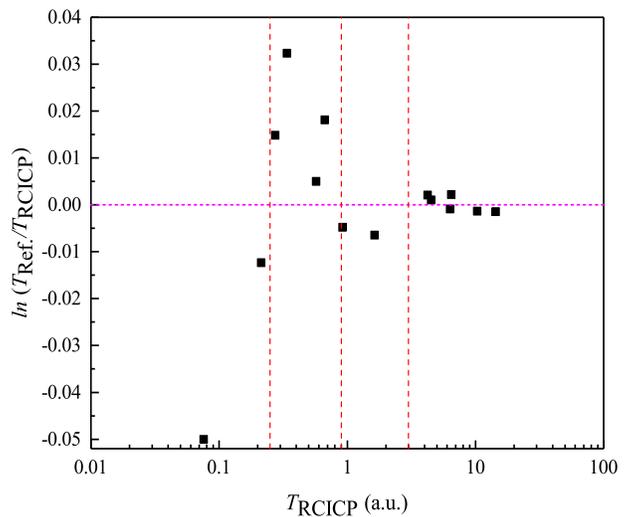}}
\vspace{0.5em}
\caption{Natural logarithm of the ratio, ln ($T_{\rm Ref}/T_{\rm RCICP}$). }
\end{figure}

\section{Remarks on uncertainty of the present results}

The uncertainties of the present transition matrix elements $T_{RCICP}$ given in Table~\ref{mat} are estimated very carefully.
Similar to the method used in Ref.~\cite{kramida2013c}, 
in which the oscillator strengths are classified into different groups based on the deviations from best experimental data,
the uncertainties of the RCICP transition matrix elements 
are evaluated by comparing with the most accurate experimental or theoretical resutls $T_{\rm Ref}$ 
listed in bold in Table \ref{mat}.
FIG.~2 plots the natural logarithm of the ratio,  ln($T_{\rm Ref}/T_{\rm RCICP}$).
As shown in this figure, the matrix elements of all the involved transitions can be classified into four groups. 
The first group is those with the $T_{\rm RCICP}$ larger than 3.0 a.u., which consists of 
the ${6s_{1/2} \rightarrow 6p_{1/2, 3/2}}$,  $7s_{1/2} \rightarrow 6p_{1/2, 3/2}$, and $7s_{1/2} \rightarrow 7p_{1/2, 3/2}$ transitions. 
These transition matrix elements are calculated very accurately.
The uncertainty estimated by the rms of the difference from the reference values is about 0.17\% for this group.
The second group is those with the $T_{\rm RCICP}$ in the range 0.9--3.0 a.u., including the ${7s_{1/2} \rightarrow 8p_{1/2,3/2}}$ transitions. 
The uncertainty for this group is about 0.6\%.
The third group is those with the $T_{\rm RCICP}$ in the range 0.25--0.9 a.u., including 
the  ${6s_{1/2} \rightarrow 7p_{1/2,3/2}}$ and  ${7s_{1/2} \rightarrow 9p_{1/2,3/2}}$ transitions. 
The uncertainty is about 2\%. 
The fourth group is those with the $T_{\rm RCICP}$ smaller than 0.25 a.u., including the ${6s_{1/2} \rightarrow 8p_{1/2,3/2}}$ transitions. 
The uncertainty is about 3.6\%.  
It should be mentioned that the uncertainties given above 
are estimated on the level of one standard deviation.

However, the uncertainties of the tune-out wavelengths need to be specially considered due to different transition contributions.
For the tune-out wavelength near 880 nm, its calculation can be simplified as
\begin{eqnarray}
\label{eqn9}
&0=\frac{f_{6s_{1/2} \rightarrow 6p_{1/2}}}{(\Delta E^2_{6s_{1/2} \rightarrow 6p_{1/2}}-\omega^2)}+
\frac{f_{6s_{1/2} \rightarrow 6p_{1/2}}\mathcal{R}_{6p}(1+\delta_{6p})}{[(\Delta E^2_{6s_{1/2} \rightarrow 6p_{1/2}})(1+\delta_{6p})^2-\omega^2]}     \nonumber  \\
& +\alpha_{rem}(\omega).
\end{eqnarray}
Here, $\mathcal{R}_{6p}$=${|\langle 6s_{1/2}||D||6p_{3/2} \rangle|^{2}}/{|\langle 6s_{1/2}||D||6p_{1/2} \rangle|^{2}}$, 
and $\alpha_{rem}(\omega)$ is the remainder part (exclude the $6s_{1/2} \rightarrow 6p_{1/2,3/2}$) of  dynamic dipole polarizability. 
The energy difference $\Delta E^2_{6p_{3/2}}$ can be expressed 
as $\Delta E^2_{6p_{3/2}}\!=\!\Delta E^2_{6p_{1/2}}$(1+$\delta_{6p})^2$ with  $\delta_{6p}$=0.0495639.
As seen from Eq.~(\ref{eqn9}), there are three sources of the uncertainty for this tune-out wavelength.
The first source is the uncertainty of the $6s_{1/2} \rightarrow 6p_{1/2}$  reduced matrix element or oscillator strengths. 
The uncertainty 0.17\% of the $6s_{1/2} \rightarrow 6p_{1/2}$ matrix elements may
lead to an uncertainty of 0.0004~nm for the 880-nm  tune-out wavelength.
The second one is the uncertainty of the line strength ratio $\mathcal{R}_{6p}$.  
We find that this tune-out wavelength is very sensitive to the ratio $\mathcal{R}_{6p}$.  
The uncertainty of the RCICP ratio $\mathcal{R}_{6p}$ = 1.9844(100) leads to an uncertainty of 0.0475~nm 
for this tune-out wavelength. 
The third one is the uncertainty of the $\alpha_{rem}(\omega)$.  
The uncertainty of the $\alpha_{rem}(\omega)$,  which includes the uncertainties of the $\alpha_{core}$ 
and transitions from more highly-excited states (assumed to be 10\% uncertainty),
may lead to an uncertainty of 0.0005~nm for this tune-out wavelength. 
By comparing the above estimations, it can be found that the total uncertainty of the 880-nm
tune-out wavelength is mainly determined by the uncertainty of the $\mathcal{R}_{6p}$, 
and the uncertainties from the uncertainties of the $6s_{1/2} \rightarrow 6p_{1/2}$ transition 
matrix element and  $\alpha_{rem}(\omega)$ are negligible.
For the 880-nm $\lambda_{\rm zero}^{recom}$, the uncertainty of the experimental 
line strength ratio ($\mathcal{R}_{6p}=1.9768(33))$ leads to an uncertainty of 0.0158 nm,
 which is about one third of the uncertainty of the RCICP tune-out wavelength.

Moreover, as for the tune-out wavelengths near 460~nm and 457~nm, 
both of them are caused by a cancellation of the $6s_{1/2} \rightarrow 6p_{1/2,3/2}$  
and $6s_{1/2} \rightarrow 7p_{1/2,3/2}$ transitions. 
The corresponding calculation can be simplified as
\begin{eqnarray}
&0=\alpha_{6p}(\omega)+\frac{f_{6s_{1/2} \rightarrow 7p_{1/2}}}
{(\Delta E^2_{7p_{1/2}}-\omega^2)}+\frac{f_{{6s_{1/2} \rightarrow 7p_{1/2}}}\mathcal{R}_{7p}
(1+\delta_{7p})}{[(\Delta E^2_{7p_{1/2}})(1+\delta_{7p})^2-\omega^2]}  \nonumber  \\
&+\alpha'_{rem}(\omega),
\end{eqnarray}
where $\mathcal{R}_{7p}={|\langle 6s_{1/2}||D||7p_{3/2} \rangle|^{2}}/{|\langle 6s_{1/2}||D||7p_{1/2} \rangle|^{2}}$. 
$\alpha'_{rem}(\omega)$ is the remainder part (exclude the $6s_{1/2} \rightarrow 6p_{1/2,3/2}$ and $6s_{1/2} \rightarrow 7p_{1/2,3/2}$ )  of dynamic dipole polarizability.
The polarizability $\alpha_{6p}(\omega)$ corresponding to the $6s_{1/2} \rightarrow 6p_{1/2,3/2}$ transitions is retained as 
a separate term since it is much larger than the remainder term. 
The uncertainty analysis of these two tune-out wavelengths is similar to 
the analysis of the 880-nm tune-out wavelength. 
The uncertainties of the RCICP $\alpha_{6p}(\omega)$ and the $6s_{1/2} \rightarrow 7p_{1/2}$ 
matrix elements lead to the uncertainties of 0.0043 nm and 0.0468 nm for these tune-out wavelengths. 
The uncertainty of the RCICP ratio $\mathcal{R}_{7p}$ = 4.3474(2463) leads to the uncertainties of 
0.0271~nm  and 0.0783~nm for 460-nm and 457-nm tune-out wavelengths, respectively. 
Consequently, the total uncertainties (arithmetic square root of the sum of squares of those uncertainties) 
of the RCICP 460-nm and 457-nm tune-out wavelength are estimated to be  about 0.0543~nm and 0.0917~nm.
However, for the 460-nm and 457-nm $\lambda_{\rm zero}^{recom}$, 
the uncertainty of the experimental $\alpha_{6p}(\omega)$, 
which were derived from the uncertainty of the experimental transition matrix elements of the $6s_{1/2} \rightarrow 6p_{1/2, 3/2}$, 
leads to an uncertainty of 0.0014~nm.  
The uncertainty of the $6s_{1/2} \rightarrow 7p_{1/2}$ matrix element makes an uncertainty of 0.0039~nm. 
The experimental ratio $\mathcal{R}_{7p}$ = 4.2626(396) leads to the uncertainties of 0.0047~nm 
and 0.0132~nm for the  460-nm and 457-nm $\lambda_{\rm zero}^{recom}$, respectively. 
As a result, the total uncertainties of the 460-nm and 457-nm $\lambda_{\rm zero}^{recom}$ are about 0.0063 nm and 0.0171nm.

Apart from the uncertainty corresponding to the fine-structure states, 
the uncertainty of tune-out wavelengths of the hyperfine states should be stated as well.
The source of the uncertainties of the tune-out wavelengths 
for the hyperfine components of the ground level should include two parts. 
The first part is from the uncertainties of the matrix elements in Table~\ref{mat}. 
The uncertainties for the primary tune-out wavelengths (near 880, 460 and 457 nm) 
derived from this part are expected to be at the same level as 
those of the tune-out wavelengths of the fine-structure $6s_{1/2}$ state.
The second part is from the uncertainties of the partial derivatives in Table~\ref{tab16}. 
This part is the main source for the uncertainties of the tune-out wavelength shifts $\Delta \lambda$ in Table~\ref{133cs}. 
In order to show the uncertainties of the $\Delta \lambda$s , Table~\ref{133cs} lists the uncertainties that are derived from the uncertainties of the partial derivatives.
However, for the tune-out wavelengths near the 894, 852, 459 and 455 nm, 
which are located between the transition energies from the hyperfine components of 
the $6p_{1/2}$, $6p_{3/2}$, $7p_{1/2}$ and $7p_{3/2}$ to the $6s_{1/2}$ $F=3$ (or $F=4$) levels, the uncertainties are estimated to be less than $10^{-7}$~nm. 

\section{CONCLUSION}

The energy levels, E1 transition matrix elements, static and dynamic polarizabilities corresponding to 
the $6s_{1/2}$ state of Cs atoms are calculated by using the RCICP approach. 
Three longest tune-out wavelengths of the $6s_{1/2}$ state are determined. 
It has been found that the present RCICP results agree very well with other available experimental and theoretical results.

Apart from the fine-structure $6s_{1/2}$ state, in particular, 
we consider the effect of the hyperfine splittings of the 
$6s_{1/2}$, 6$p_j$, and 7$p_{j}$ states of $^{133}$Cs atoms. 
The dynamic polarizabilities and tune-out wavelengths of the hyperfine components of the ground level of $^{133}$Cs atoms are further determined. 
The tune-out wavelengths corresponding to the hyperfine splittings of the $np_j$ states are hard to be measured. 
However, all of the obtained primary tune-out wavelengths have relatively large hyperfine shifts. 
For example, the shifts of the 880-nm tune-out wavelength corresponding to the 6$s_{1/2}, F=3$ and $F=4$ states 
are about $-$0.0135 nm and 0.0106 nm, respectively; they are very close to the hyperfine interaction energies. 
Such large shifts can be resolved in experiments. If the tune-out wavelengths could be measured at a level of fm, 
the contribution of the tensor polarizabilities of the hyperfine  
components of the ground level to the tune-out wavelengths could be extracted from experiments. 
In addition, the 880-nm tune-out wavelength is found to be very sensitive to the ratio of the line strengths of the $6s_{1/2} \to  6p_{1/2,3/2}$ transitions. 
Accurate measurements on this tune-out wavelength would enable a high-precision determination of the $6s_{1/2} \to  6p_{1/2, 3/2}$  line strength ratio.

\section{ACKNOWLEDGMENTS}

This work has been supported by the National Key Research and Development Program of China under Grant No.~2017YFA0402300 and the National Natural Science Foundation of China under Grant Nos. 11774292, 11564036, 11804280,
11874051, and 11864036. Z.~W.~W. acknowledges the Major Project of the Research Ability Promotion Program for Young Scholars of Northwest Normal University of China under Grant No.~NWNU-LKQN2019-5. We are very grateful to the anonymous referees for their constructive comments on this work.


\begin{thebibliography}{94}%
\makeatletter
\providecommand \@ifxundefined [1]{%
 \@ifx{#1\undefined}
}%
\providecommand \@ifnum [1]{%
 \ifnum #1\expandafter \@firstoftwo
 \else \expandafter \@secondoftwo
 \fi
}%
\providecommand \@ifx [1]{%
 \ifx #1\expandafter \@firstoftwo
 \else \expandafter \@secondoftwo
 \fi
}%
\providecommand \natexlab [1]{#1}%
\providecommand \enquote  [1]{``#1''}%
\providecommand \bibnamefont  [1]{#1}%
\providecommand \bibfnamefont [1]{#1}%
\providecommand \citenamefont [1]{#1}%
\providecommand \href@noop [0]{\@secondoftwo}%
\providecommand \href [0]{\begingroup \@sanitize@url \@href}%
\providecommand \@href[1]{\@@startlink{#1}\@@href}%
\providecommand \@@href[1]{\endgroup#1\@@endlink}%
\providecommand \@sanitize@url [0]{\catcode `\\12\catcode `\$12\catcode
  `\&12\catcode `\#12\catcode `\^12\catcode `\_12\catcode `\%12\relax}%
\providecommand \@@startlink[1]{}%
\providecommand \@@endlink[0]{}%
\providecommand \url  [0]{\begingroup\@sanitize@url \@url }%
\providecommand \@url [1]{\endgroup\@href {#1}{\urlprefix }}%
\providecommand \urlprefix  [0]{URL }%
\providecommand \Eprint [0]{\href }%
\providecommand \doibase [0]{http://dx.doi.org/}%
\providecommand \selectlanguage [0]{\@gobble}%
\providecommand \bibinfo  [0]{\@secondoftwo}%
\providecommand \bibfield  [0]{\@secondoftwo}%
\providecommand \translation [1]{[#1]}%
\providecommand \BibitemOpen [0]{}%
\providecommand \bibitemStop [0]{}%
\providecommand \bibitemNoStop [0]{.\EOS\space}%
\providecommand \EOS [0]{\spacefactor3000\relax}%
\providecommand \BibitemShut  [1]{\csname bibitem#1\endcsname}%
\let\auto@bib@innerbib\@empty
\bibitem [{\citenamefont {{Grimm}}\ \emph {et~al.}(2000)\citenamefont
  {{Grimm}}, \citenamefont {{Weidem{\"u}ller}},\ and\ \citenamefont
  {{Ovchinnikov}}}]{Grimm2000}%
  \BibitemOpen
  \bibfield  {author} {\bibinfo {author} {\bibfnamefont {R.}~\bibnamefont
  {{Grimm}}}, \bibinfo {author} {\bibfnamefont {M.}~\bibnamefont
  {{Weidem{\"u}ller}}}, \ and\ \bibinfo {author} {\bibfnamefont {Y.~B.}\
  \bibnamefont {{Ovchinnikov}}},\ }\href {\doibase
  10.1016/S1049-250X(08)60186-X} {\bibfield  {journal} {\bibinfo  {journal}
  {Adv.~At.~Mol.~Opt.~Phys.}\ }\textbf {\bibinfo {volume} {42}},\ \bibinfo
  {pages} {95} (\bibinfo {year} {2000})}\BibitemShut {NoStop}%
\bibitem [{\citenamefont {{Leblanc}}\ and\ \citenamefont
  {{Thywissen}}(2007)}]{LeBlanc2007}%
  \BibitemOpen
  \bibfield  {author} {\bibinfo {author} {\bibfnamefont {L.~J.}\ \bibnamefont
  {{Leblanc}}}\ and\ \bibinfo {author} {\bibfnamefont {J.~H.}\ \bibnamefont
  {{Thywissen}}},\ }\href {\doibase 10.1103/PhysRevA.75.053612} {\bibfield
  {journal} {\bibinfo  {journal} {\pra}\ }\textbf {\bibinfo {volume} {75}},\
  \bibinfo {eid} {053612} (\bibinfo {year} {2007})}\BibitemShut {NoStop}%
\bibitem [{\citenamefont {{Luan}}\ \emph {et~al.}(2015)\citenamefont {{Luan}},
  \citenamefont {{Yao}}, \citenamefont {{Wang}}, \citenamefont {{Li}},
  \citenamefont {{Yang}}, \citenamefont {{Chen}},\ and\ \citenamefont
  {{Ma}}}]{Luan2015}%
  \BibitemOpen
  \bibfield  {author} {\bibinfo {author} {\bibfnamefont {T.}~\bibnamefont
  {{Luan}}}, \bibinfo {author} {\bibfnamefont {H.}~\bibnamefont {{Yao}}},
  \bibinfo {author} {\bibfnamefont {L.}~\bibnamefont {{Wang}}}, \bibinfo
  {author} {\bibfnamefont {C.}~\bibnamefont {{Li}}}, \bibinfo {author}
  {\bibfnamefont {S.}~\bibnamefont {{Yang}}}, \bibinfo {author} {\bibfnamefont
  {X.}~\bibnamefont {{Chen}}}, \ and\ \bibinfo {author} {\bibfnamefont
  {Z.}~\bibnamefont {{Ma}}},\ }\href {\doibase 10.1364/OE.23.011378} {\bibfield
   {journal} {\bibinfo  {journal} {Opt. Express}\ }\textbf {\bibinfo {volume}
  {23}},\ \bibinfo {pages} {11378} (\bibinfo {year} {2015})}\BibitemShut
  {NoStop}%
\bibitem [{\citenamefont {{Chamakhi}}\ \emph {et~al.}(2015)\citenamefont
  {{Chamakhi}}, \citenamefont {{Ahlers}}, \citenamefont {{Telmini}},
  \citenamefont {{Schubert}}, \citenamefont {{Rasel}},\ and\ \citenamefont
  {{Gaaloul}}}]{Chamakhi2015}%
  \BibitemOpen
  \bibfield  {author} {\bibinfo {author} {\bibfnamefont {R.}~\bibnamefont
  {{Chamakhi}}}, \bibinfo {author} {\bibfnamefont {H.}~\bibnamefont
  {{Ahlers}}}, \bibinfo {author} {\bibfnamefont {M.}~\bibnamefont {{Telmini}}},
  \bibinfo {author} {\bibfnamefont {C.}~\bibnamefont {{Schubert}}}, \bibinfo
  {author} {\bibfnamefont {E.~M.}\ \bibnamefont {{Rasel}}}, \ and\ \bibinfo
  {author} {\bibfnamefont {N.}~\bibnamefont {{Gaaloul}}},\ }\href {\doibase
  10.1088/1367-2630/17/12/123002} {\bibfield  {journal} {\bibinfo  {journal}
  {New J. Phys.}\ }\textbf {\bibinfo {volume} {17}},\ \bibinfo {eid} {123002}
  (\bibinfo {year} {2015})}\BibitemShut {NoStop}%
\bibitem [{\citenamefont {{Catani}}\ \emph {et~al.}(2009)\citenamefont
  {{Catani}}, \citenamefont {{Barontini}}, \citenamefont {{Lamporesi}},
  \citenamefont {{Rabatti}}, \citenamefont {{Thalhammer}}, \citenamefont
  {{Minardi}}, \citenamefont {{Stringari}},\ and\ \citenamefont
  {{Inguscio}}}]{Catani2009}%
  \BibitemOpen
  \bibfield  {author} {\bibinfo {author} {\bibfnamefont {J.}~\bibnamefont
  {{Catani}}}, \bibinfo {author} {\bibfnamefont {G.}~\bibnamefont
  {{Barontini}}}, \bibinfo {author} {\bibfnamefont {G.}~\bibnamefont
  {{Lamporesi}}}, \bibinfo {author} {\bibfnamefont {F.}~\bibnamefont
  {{Rabatti}}}, \bibinfo {author} {\bibfnamefont {G.}~\bibnamefont
  {{Thalhammer}}}, \bibinfo {author} {\bibfnamefont {F.}~\bibnamefont
  {{Minardi}}}, \bibinfo {author} {\bibfnamefont {S.}~\bibnamefont
  {{Stringari}}}, \ and\ \bibinfo {author} {\bibfnamefont {M.}~\bibnamefont
  {{Inguscio}}},\ }\href {\doibase 10.1103/PhysRevLett.103.140401} {\bibfield
  {journal} {\bibinfo  {journal} {\prl}\ }\textbf {\bibinfo {volume} {103}},\
  \bibinfo {eid} {140401} (\bibinfo {year} {2009})}\BibitemShut {NoStop}%
\bibitem [{\citenamefont {{Herold}}\ \emph {et~al.}(2012)\citenamefont
  {{Herold}}, \citenamefont {{Vaidya}}, \citenamefont {{Li}}, \citenamefont
  {{Rolston}}, \citenamefont {{Porto}},\ and\ \citenamefont
  {{Safronova}}}]{Herold2012}%
  \BibitemOpen
  \bibfield  {author} {\bibinfo {author} {\bibfnamefont {C.~D.}\ \bibnamefont
  {{Herold}}}, \bibinfo {author} {\bibfnamefont {V.~D.}\ \bibnamefont
  {{Vaidya}}}, \bibinfo {author} {\bibfnamefont {X.}~\bibnamefont {{Li}}},
  \bibinfo {author} {\bibfnamefont {S.~L.}\ \bibnamefont {{Rolston}}}, \bibinfo
  {author} {\bibfnamefont {J.~V.}\ \bibnamefont {{Porto}}}, \ and\ \bibinfo
  {author} {\bibfnamefont {M.~S.}\ \bibnamefont {{Safronova}}},\ }\href
  {\doibase 10.1103/PhysRevLett.109.243003} {\bibfield  {journal} {\bibinfo
  {journal} {\prl}\ }\textbf {\bibinfo {volume} {109}},\ \bibinfo {eid}
  {243003} (\bibinfo {year} {2012})}\BibitemShut {NoStop}%
\bibitem [{\citenamefont {{Safronova}}\ \emph {et~al.}(2015)\citenamefont
  {{Safronova}}, \citenamefont {{Zuhrianda}}, \citenamefont {{Safronova}},\
  and\ \citenamefont {{Clark}}}]{Safronova2015}%
  \BibitemOpen
  \bibfield  {author} {\bibinfo {author} {\bibfnamefont {M.~S.}\ \bibnamefont
  {{Safronova}}}, \bibinfo {author} {\bibfnamefont {Z.}~\bibnamefont
  {{Zuhrianda}}}, \bibinfo {author} {\bibfnamefont {U.~I.}\ \bibnamefont
  {{Safronova}}}, \ and\ \bibinfo {author} {\bibfnamefont {C.~W.}\ \bibnamefont
  {{Clark}}},\ }\href {\doibase 10.1103/PhysRevA.92.040501} {\bibfield
  {journal} {\bibinfo  {journal} {\pra}\ }\textbf {\bibinfo {volume} {92}},\
  \bibinfo {eid} {040501} (\bibinfo {year} {2015})}\BibitemShut {NoStop}%
\bibitem [{\citenamefont {{Fallon}}\ and\ \citenamefont
  {{Sackett}}(2016)}]{Fallon2016}%
  \BibitemOpen
  \bibfield  {author} {\bibinfo {author} {\bibfnamefont {A.}~\bibnamefont
  {{Fallon}}}\ and\ \bibinfo {author} {\bibfnamefont {C.}~\bibnamefont
  {{Sackett}}},\ }\href {\doibase 10.3390/atoms4020012} {\bibfield  {journal}
  {\bibinfo  {journal} {Atoms}\ }\textbf {\bibinfo {volume} {4}},\ \bibinfo
  {pages} {12} (\bibinfo {year} {2016})}\BibitemShut {NoStop}%
\bibitem [{\citenamefont {{Holmgren}}\ \emph {et~al.}(2012)\citenamefont
  {{Holmgren}}, \citenamefont {{Trubko}}, \citenamefont {{Hromada}},\ and\
  \citenamefont {{Cronin}}}]{Holmgren2012}%
  \BibitemOpen
  \bibfield  {author} {\bibinfo {author} {\bibfnamefont {W.~F.}\ \bibnamefont
  {{Holmgren}}}, \bibinfo {author} {\bibfnamefont {R.}~\bibnamefont
  {{Trubko}}}, \bibinfo {author} {\bibfnamefont {I.}~\bibnamefont {{Hromada}}},
  \ and\ \bibinfo {author} {\bibfnamefont {A.~D.}\ \bibnamefont {{Cronin}}},\
  }\href {\doibase 10.1103/PhysRevLett.109.243004} {\bibfield  {journal}
  {\bibinfo  {journal} {\prl}\ }\textbf {\bibinfo {volume} {109}},\ \bibinfo
  {eid} {243004} (\bibinfo {year} {2012})}\BibitemShut {NoStop}%
\bibitem [{\citenamefont {{Leonard}}\ \emph {et~al.}(2015)\citenamefont
  {{Leonard}}, \citenamefont {{Fallon}}, \citenamefont {{Sackett}},\ and\
  \citenamefont {{Safronova}}}]{Leonard2015}%
  \BibitemOpen
  \bibfield  {author} {\bibinfo {author} {\bibfnamefont {R.~H.}\ \bibnamefont
  {{Leonard}}}, \bibinfo {author} {\bibfnamefont {A.~J.}\ \bibnamefont
  {{Fallon}}}, \bibinfo {author} {\bibfnamefont {C.~A.}\ \bibnamefont
  {{Sackett}}}, \ and\ \bibinfo {author} {\bibfnamefont {M.~S.}\ \bibnamefont
  {{Safronova}}},\ }\href {\doibase 10.1103/PhysRevA.92.052501} {\bibfield
  {journal} {\bibinfo  {journal} {\pra}\ }\textbf {\bibinfo {volume} {92}},\
  \bibinfo {eid} {052501} (\bibinfo {year} {2015})}\BibitemShut {NoStop}%
\bibitem [{\citenamefont {{Leonard}}\ \emph {et~al.}(2017)\citenamefont
  {{Leonard}}, \citenamefont {{Fallon}}, \citenamefont {{Sackett}},\ and\
  \citenamefont {{Safronova}}}]{Leonard2017}%
  \BibitemOpen
  \bibfield  {author} {\bibinfo {author} {\bibfnamefont {R.~H.}\ \bibnamefont
  {{Leonard}}}, \bibinfo {author} {\bibfnamefont {A.~J.}\ \bibnamefont
  {{Fallon}}}, \bibinfo {author} {\bibfnamefont {C.~A.}\ \bibnamefont
  {{Sackett}}}, \ and\ \bibinfo {author} {\bibfnamefont {M.~S.}\ \bibnamefont
  {{Safronova}}},\ }\href {\doibase 10.1103/PhysRevA.95.059901} {\bibfield
  {journal} {\bibinfo  {journal} {\pra}\ }\textbf {\bibinfo {volume} {95}},\
  \bibinfo {eid} {059901} (\bibinfo {year} {2017})}\BibitemShut {NoStop}%
\bibitem [{\citenamefont {{Henson}}\ \emph {et~al.}(2015)\citenamefont
  {{Henson}}, \citenamefont {{Khakimov}}, \citenamefont {{Dall}}, \citenamefont
  {{Baldwin}}, \citenamefont {{Tang}},\ and\ \citenamefont
  {{Truscott}}}]{Henson2015}%
  \BibitemOpen
  \bibfield  {author} {\bibinfo {author} {\bibfnamefont {B.~M.}\ \bibnamefont
  {{Henson}}}, \bibinfo {author} {\bibfnamefont {R.~I.}\ \bibnamefont
  {{Khakimov}}}, \bibinfo {author} {\bibfnamefont {R.~G.}\ \bibnamefont
  {{Dall}}}, \bibinfo {author} {\bibfnamefont {K.~G.~H.}\ \bibnamefont
  {{Baldwin}}}, \bibinfo {author} {\bibfnamefont {L.-Y.}\ \bibnamefont
  {{Tang}}}, \ and\ \bibinfo {author} {\bibfnamefont {A.~G.}\ \bibnamefont
  {{Truscott}}},\ }\href {\doibase 10.1103/PhysRevLett.115.043004} {\bibfield
  {journal} {\bibinfo  {journal} {\prl}\ }\textbf {\bibinfo {volume} {115}},\
  \bibinfo {eid} {043004} (\bibinfo {year} {2015})}\BibitemShut {NoStop}%
\bibitem [{\citenamefont {{Mitroy}}\ and\ \citenamefont
  {{Tang}}(2013)}]{Mitroy2013}%
  \BibitemOpen
  \bibfield  {author} {\bibinfo {author} {\bibfnamefont {J.}~\bibnamefont
  {{Mitroy}}}\ and\ \bibinfo {author} {\bibfnamefont {L.-Y.}\ \bibnamefont
  {{Tang}}},\ }\href {\doibase 10.1103/PhysRevA.88.052515} {\bibfield
  {journal} {\bibinfo  {journal} {\pra}\ }\textbf {\bibinfo {volume} {88}},\
  \bibinfo {eid} {052515} (\bibinfo {year} {2013})}\BibitemShut {NoStop}%
\bibitem [{\citenamefont {{Copenhaver}}\ \emph {et~al.}(2019)\citenamefont
  {{Copenhaver}}, \citenamefont {{Cassella}}, \citenamefont {{Berghaus}},\ and\
  \citenamefont {{M{\"u}ller}}}]{Copenhaver2019}%
  \BibitemOpen
  \bibfield  {author} {\bibinfo {author} {\bibfnamefont {E.}~\bibnamefont
  {{Copenhaver}}}, \bibinfo {author} {\bibfnamefont {K.}~\bibnamefont
  {{Cassella}}}, \bibinfo {author} {\bibfnamefont {R.}~\bibnamefont
  {{Berghaus}}}, \ and\ \bibinfo {author} {\bibfnamefont {H.}~\bibnamefont
  {{M{\"u}ller}}},\ }\href {\doibase 10.1103/PhysRevA.100.063603} {\bibfield
  {journal} {\bibinfo  {journal} {\pra}\ }\textbf {\bibinfo {volume} {100}},\
  \bibinfo {eid} {063603} (\bibinfo {year} {2019})}\BibitemShut {NoStop}%
\bibitem [{\citenamefont {{Monroe}}\ \emph {et~al.}(1995)\citenamefont
  {{Monroe}}, \citenamefont {{Meekhof}}, \citenamefont {{King}}, \citenamefont
  {{Itano}},\ and\ \citenamefont {{Wineland}}}]{Demonstration1995}%
  \BibitemOpen
  \bibfield  {author} {\bibinfo {author} {\bibfnamefont {C.}~\bibnamefont
  {{Monroe}}}, \bibinfo {author} {\bibfnamefont {D.~M.}\ \bibnamefont
  {{Meekhof}}}, \bibinfo {author} {\bibfnamefont {B.~E.}\ \bibnamefont
  {{King}}}, \bibinfo {author} {\bibfnamefont {W.~M.}\ \bibnamefont {{Itano}}},
  \ and\ \bibinfo {author} {\bibfnamefont {D.~J.}\ \bibnamefont {{Wineland}}},\
  }\href {\doibase 10.1103/PhysRevLett.75.4714} {\bibfield  {journal} {\bibinfo
   {journal} {\prl}\ }\textbf {\bibinfo {volume} {75}},\ \bibinfo {pages}
  {4714} (\bibinfo {year} {1995})}\BibitemShut {NoStop}%
\bibitem [{\citenamefont {Jefferts}\ \emph {et~al.}(2014)\citenamefont
  {Jefferts}, \citenamefont {Heavner}, \citenamefont {Parker}, \citenamefont
  {Shirley}, \citenamefont {Donley}, \citenamefont {Ashby}, \citenamefont
  {Levi}, \citenamefont {Calonico},\ and\ \citenamefont
  {Costanzo}}]{Jefferts2014}%
  \BibitemOpen
  \bibfield  {author} {\bibinfo {author} {\bibfnamefont {S.~R.}\ \bibnamefont
  {Jefferts}}, \bibinfo {author} {\bibfnamefont {T.~P.}\ \bibnamefont
  {Heavner}}, \bibinfo {author} {\bibfnamefont {T.~E.}\ \bibnamefont {Parker}},
  \bibinfo {author} {\bibfnamefont {J.~H.}\ \bibnamefont {Shirley}}, \bibinfo
  {author} {\bibfnamefont {E.~A.}\ \bibnamefont {Donley}}, \bibinfo {author}
  {\bibfnamefont {N.}~\bibnamefont {Ashby}}, \bibinfo {author} {\bibfnamefont
  {F.}~\bibnamefont {Levi}}, \bibinfo {author} {\bibfnamefont {D.}~\bibnamefont
  {Calonico}}, \ and\ \bibinfo {author} {\bibfnamefont {G.~A.}\ \bibnamefont
  {Costanzo}},\ }\href {\doibase 10.1103/PhysRevLett.112.050801} {\bibfield
  {journal} {\bibinfo  {journal} {Phys. Rev. Lett.}\ }\textbf {\bibinfo
  {volume} {112}},\ \bibinfo {pages} {050801} (\bibinfo {year}
  {2014})}\BibitemShut {NoStop}%
\bibitem [{\citenamefont {{Wang}}\ \emph {et~al.}(2015)\citenamefont {{Wang}},
  \citenamefont {{Zhang}}, \citenamefont {{Corcovilos}}, \citenamefont
  {{Kumar}},\ and\ \citenamefont {{Weiss}}}]{WangYang2015}%
  \BibitemOpen
  \bibfield  {author} {\bibinfo {author} {\bibfnamefont {Y.}~\bibnamefont
  {{Wang}}}, \bibinfo {author} {\bibfnamefont {X.}~\bibnamefont {{Zhang}}},
  \bibinfo {author} {\bibfnamefont {T.~A.}\ \bibnamefont {{Corcovilos}}},
  \bibinfo {author} {\bibfnamefont {A.}~\bibnamefont {{Kumar}}}, \ and\
  \bibinfo {author} {\bibfnamefont {D.~S.}\ \bibnamefont {{Weiss}}},\ }\href
  {\doibase 10.1103/PhysRevLett.115.043003} {\bibfield  {journal} {\bibinfo
  {journal} {\prl}\ }\textbf {\bibinfo {volume} {115}},\ \bibinfo {eid}
  {043003} (\bibinfo {year} {2015})}\BibitemShut {NoStop}%
\bibitem [{\citenamefont {{Arora}}\ \emph {et~al.}(2011)\citenamefont
  {{Arora}}, \citenamefont {{Safronova}},\ and\ \citenamefont
  {{Clark}}}]{Arora2011}%
  \BibitemOpen
  \bibfield  {author} {\bibinfo {author} {\bibfnamefont {B.}~\bibnamefont
  {{Arora}}}, \bibinfo {author} {\bibfnamefont {M.~S.}\ \bibnamefont
  {{Safronova}}}, \ and\ \bibinfo {author} {\bibfnamefont {C.~W.}\ \bibnamefont
  {{Clark}}},\ }\href {\doibase 10.1103/PhysRevA.84.043401} {\bibfield
  {journal} {\bibinfo  {journal} {\pra}\ }\textbf {\bibinfo {volume} {84}},\
  \bibinfo {eid} {043401} (\bibinfo {year} {2011})}\BibitemShut {NoStop}%
\bibitem [{\citenamefont {Mitroy}\ and\ \citenamefont
  {Bromley}(2003)}]{Mitroy2003Semiempirical}%
  \BibitemOpen
  \bibfield  {author} {\bibinfo {author} {\bibfnamefont {J.}~\bibnamefont
  {Mitroy}}\ and\ \bibinfo {author} {\bibfnamefont {M.~W.~J.}\ \bibnamefont
  {Bromley}},\ }\href@noop {} {\bibfield  {journal} {\bibinfo  {journal} {Phys.
  Rev. A}\ }\textbf {\bibinfo {volume} {68}},\ \bibinfo {pages} {52714}
  (\bibinfo {year} {2003})}\BibitemShut {NoStop}%
\bibitem [{\citenamefont {Grant}(2007)}]{Grant2007}%
  \BibitemOpen
  \bibfield  {author} {\bibinfo {author} {\bibfnamefont {I.~P.}\ \bibnamefont
  {Grant}},\ }\href@noop {} {\emph {\bibinfo {title} {Relativistic Quantum
  Theory of Atoms and Molecules Theory and Computation}}}\ (\bibinfo
  {publisher} {Springer},\ \bibinfo {address} {New York},\ \bibinfo {year}
  {2007})\BibitemShut {NoStop}%
\bibitem [{\citenamefont {Grant}\ and\ \citenamefont
  {Quiney}(1988)}]{grant1988}%
  \BibitemOpen
  \bibfield  {author} {\bibinfo {author} {\bibfnamefont {I.~P.}\ \bibnamefont
  {Grant}}\ and\ \bibinfo {author} {\bibfnamefont {H.~M.}\ \bibnamefont
  {Quiney}},\ }\href@noop {} {\bibfield  {journal} {\bibinfo  {journal} {Adv.
  At. Mol. Phys}\ }\textbf {\bibinfo {volume} {23}},\ \bibinfo {pages} {37}
  (\bibinfo {year} {1988})}\BibitemShut {NoStop}%
\bibitem [{\citenamefont {Grant}(1989)}]{grant1989}%
  \BibitemOpen
  \bibfield  {author} {\bibinfo {author} {\bibfnamefont {I.~P.}\ \bibnamefont
  {Grant}},\ }\href@noop {} {\bibfield  {journal} {\bibinfo  {journal}
  {Relativistic, Quantum Electrodynamic and Weak Interaction Effects in Atoms,
  edited by P.J.Mohr, W. R. Johnson, and J. Sucher, AIP Conf. Proc. No. 189
  (AIP, New York)}\ ,\ \bibinfo {pages} {pp. 235}} (\bibinfo {year}
  {1989})}\BibitemShut {NoStop}%
\bibitem [{\citenamefont {Grant}(1996)}]{grant1996}%
  \BibitemOpen
  \bibfield  {author} {\bibinfo {author} {\bibfnamefont {I.~P.}\ \bibnamefont
  {Grant}},\ }\href@noop {} {\bibfield  {journal} {\bibinfo  {journal} {Advance
  Atom, Molecular and Optical Physics Reference Book, edited by G. W. F. Drake
  (AIP, New York}\ }\textbf {\bibinfo {volume} {Chap. 22}} (\bibinfo {year}
  {1996})}\BibitemShut {NoStop}%
\bibitem [{\citenamefont {Lim}\ \emph {et~al.}(2002)\citenamefont {Lim},
  \citenamefont {Laerdahl},\ and\ \citenamefont {Schwerdtfeger}}]{Lim2002}%
  \BibitemOpen
  \bibfield  {author} {\bibinfo {author} {\bibfnamefont {I.}~\bibnamefont
  {Lim}}, \bibinfo {author} {\bibfnamefont {J.}~\bibnamefont {Laerdahl}}, \
  and\ \bibinfo {author} {\bibfnamefont {P.}~\bibnamefont {Schwerdtfeger}},\
  }\href {\doibase 10.1063/1.1420747} {\bibfield  {journal} {\bibinfo
  {journal} {J. Chem. Phys.}\ }\textbf {\bibinfo {volume} {116}},\ \bibinfo
  {pages} {172} (\bibinfo {year} {2002})}\BibitemShut {NoStop}%
\bibitem [{\citenamefont {{Johnson}}\ \emph {et~al.}(1983)\citenamefont
  {{Johnson}}, \citenamefont {{Kolb}},\ and\ \citenamefont
  {{Huang}}}]{Johnson1983}%
  \BibitemOpen
  \bibfield  {author} {\bibinfo {author} {\bibfnamefont {W.~R.}\ \bibnamefont
  {{Johnson}}}, \bibinfo {author} {\bibfnamefont {D.}~\bibnamefont {{Kolb}}}, \
  and\ \bibinfo {author} {\bibfnamefont {K.~N.}\ \bibnamefont {{Huang}}},\
  }\href {\doibase 10.1016/0092-640X(83)90020-7} {\bibfield  {journal}
  {\bibinfo  {journal} {At.~Data~Nucl.~Data~Tables}\ }\textbf {\bibinfo
  {volume} {28}},\ \bibinfo {pages} {333} (\bibinfo {year} {1983})}\BibitemShut
  {NoStop}%
\bibitem [{\citenamefont {{Grant}}\ and\ \citenamefont
  {{Quiney}}(2000)}]{Grant2000}%
  \BibitemOpen
  \bibfield  {author} {\bibinfo {author} {\bibfnamefont {I.~P.}\ \bibnamefont
  {{Grant}}}\ and\ \bibinfo {author} {\bibfnamefont {H.~M.}\ \bibnamefont
  {{Quiney}}},\ }\href {\doibase 10.1103/PhysRevA.62.022508} {\bibfield
  {journal} {\bibinfo  {journal} {\pra}\ }\textbf {\bibinfo {volume} {62}},\
  \bibinfo {eid} {022508} (\bibinfo {year} {2000})}\BibitemShut {NoStop}%
\bibitem [{\citenamefont {{Mitroy}}\ and\ \citenamefont
  {{Bromley}}(2003)}]{Mitroy2003}%
  \BibitemOpen
  \bibfield  {author} {\bibinfo {author} {\bibfnamefont {J.}~\bibnamefont
  {{Mitroy}}}\ and\ \bibinfo {author} {\bibfnamefont {M.~W.}\ \bibnamefont
  {{Bromley}}},\ }\href {\doibase 10.1103/PhysRevA.68.052714} {\bibfield
  {journal} {\bibinfo  {journal} {\pra}\ }\textbf {\bibinfo {volume} {68}},\
  \bibinfo {eid} {052714} (\bibinfo {year} {2003})}\BibitemShut {NoStop}%
\bibitem [{\citenamefont {Kramida}\ \emph {et~al.}()\citenamefont {Kramida},
  \citenamefont {{Yu.~Ralchenko}}, \citenamefont {Reader},\ and\ \citenamefont
  {{and NIST ASD Team}}}]{Kramida2015}%
  \BibitemOpen
  \bibfield  {author} {\bibinfo {author} {\bibfnamefont {A.}~\bibnamefont
  {Kramida}}, \bibinfo {author} {\bibnamefont {{Yu.~Ralchenko}}}, \bibinfo
  {author} {\bibfnamefont {J.}~\bibnamefont {Reader}}, \ and\ \bibinfo {author}
  {\bibnamefont {{and NIST ASD Team}}},\ }\href@noop {} {}\bibinfo
  {howpublished} {NIST Atomic Spectra Database (Ver. 5.7.1),2019, [Online].
  Available: {\tt{https://physics.nist.gov/asd}}}\BibitemShut {NoStop}%
\bibitem [{\citenamefont {{Safronova}}\ \emph {et~al.}()\citenamefont
  {{Safronova}}, \citenamefont {{Safronova}},\ and\ \citenamefont
  {{Clark}}}]{Safronova2016}%
  \BibitemOpen
  \bibfield  {author} {\bibinfo {author} {\bibfnamefont {M.~S.}\ \bibnamefont
  {{Safronova}}}, \bibinfo {author} {\bibfnamefont {U.~I.}\ \bibnamefont
  {{Safronova}}}, \ and\ \bibinfo {author} {\bibfnamefont {C.~W.}\ \bibnamefont
  {{Clark}}},\ }\href@noop {} {\bibinfo  {journal} {Phys. Rev. A 94, 012505
  (2016), and Supplement Material
  https://journals.aps.org/pra/supplemental/0.1103/
  PhysRevA.94.012505/Suppl.pdf}\ }\BibitemShut {NoStop}%
\bibitem [{\citenamefont {{Safronova}}\ \emph {et~al.}(1999)\citenamefont
  {{Safronova}}, \citenamefont {{Johnson}},\ and\ \citenamefont
  {{Derevianko}}}]{Safronova1999}%
  \BibitemOpen
\bibfield  {journal} {  }\bibfield  {author} {\bibinfo {author} {\bibfnamefont
  {M.~S.}\ \bibnamefont {{Safronova}}}, \bibinfo {author} {\bibfnamefont
  {W.~R.}\ \bibnamefont {{Johnson}}}, \ and\ \bibinfo {author} {\bibfnamefont
  {A.}~\bibnamefont {{Derevianko}}},\ }\href {\doibase
  10.1103/PhysRevA.60.4476} {\bibfield  {journal} {\bibinfo  {journal} {\pra}\
  }\textbf {\bibinfo {volume} {60}},\ \bibinfo {pages} {4476} (\bibinfo {year}
  {1999})}\BibitemShut {NoStop}%
\bibitem [{\citenamefont {{Roberts}}\ \emph {et~al.}(2014)\citenamefont
  {{Roberts}}, \citenamefont {{Dzuba}},\ and\ \citenamefont
  {{Flambaum}}}]{Roberts2014}%
  \BibitemOpen
  \bibfield  {author} {\bibinfo {author} {\bibfnamefont {B.~M.}\ \bibnamefont
  {{Roberts}}}, \bibinfo {author} {\bibfnamefont {V.~A.}\ \bibnamefont
  {{Dzuba}}}, \ and\ \bibinfo {author} {\bibfnamefont {V.~V.}\ \bibnamefont
  {{Flambaum}}},\ }\href {\doibase 10.1103/PhysRevA.89.012502} {\bibfield
  {journal} {\bibinfo  {journal} {\pra}\ }\textbf {\bibinfo {volume} {89}},\
  \bibinfo {eid} {012502} (\bibinfo {year} {2014})}\BibitemShut {NoStop}%
\bibitem [{\citenamefont {{Rafac}}\ \emph {et~al.}(1999)\citenamefont
  {{Rafac}}, \citenamefont {{Tanner}}, \citenamefont {{Livingston}},\ and\
  \citenamefont {{Berry}}}]{Rafac1999}%
  \BibitemOpen
  \bibfield  {author} {\bibinfo {author} {\bibfnamefont {R.~J.}\ \bibnamefont
  {{Rafac}}}, \bibinfo {author} {\bibfnamefont {C.~E.}\ \bibnamefont
  {{Tanner}}}, \bibinfo {author} {\bibfnamefont {A.~E.}\ \bibnamefont
  {{Livingston}}}, \ and\ \bibinfo {author} {\bibfnamefont {H.~G.}\
  \bibnamefont {{Berry}}},\ }\href {\doibase 10.1103/PhysRevA.60.3648}
  {\bibfield  {journal} {\bibinfo  {journal} {\pra}\ }\textbf {\bibinfo
  {volume} {60}},\ \bibinfo {pages} {3648} (\bibinfo {year}
  {1999})}\BibitemShut {NoStop}%
\bibitem [{\citenamefont {{Derevianko}}\ and\ \citenamefont
  {{Porsev}}(2002)}]{Derevianko2002}%
  \BibitemOpen
  \bibfield  {author} {\bibinfo {author} {\bibfnamefont {A.}~\bibnamefont
  {{Derevianko}}}\ and\ \bibinfo {author} {\bibfnamefont {S.~G.}\ \bibnamefont
  {{Porsev}}},\ }\href {\doibase 10.1103/PhysRevA.65.053403} {\bibfield
  {journal} {\bibinfo  {journal} {\pra}\ }\textbf {\bibinfo {volume} {65}},\
  \bibinfo {pages} {053403} (\bibinfo {year} {2002})}\BibitemShut {NoStop}%
\bibitem [{\citenamefont {{Gregoire}}\ \emph {et~al.}(2016)\citenamefont
  {{Gregoire}}, \citenamefont {{Brooks}}, \citenamefont {{Trubko}},\ and\
  \citenamefont {{Cronin}}}]{Gregoire2016}%
  \BibitemOpen
  \bibfield  {author} {\bibinfo {author} {\bibfnamefont {M.}~\bibnamefont
  {{Gregoire}}}, \bibinfo {author} {\bibfnamefont {N.}~\bibnamefont
  {{Brooks}}}, \bibinfo {author} {\bibfnamefont {R.}~\bibnamefont {{Trubko}}},
  \ and\ \bibinfo {author} {\bibfnamefont {A.}~\bibnamefont {{Cronin}}},\
  }\href {\doibase 10.3390/atoms4030021} {\bibfield  {journal} {\bibinfo
  {journal} {Atoms}\ }\textbf {\bibinfo {volume} {4}},\ \bibinfo {pages} {21}
  (\bibinfo {year} {2016})}\BibitemShut {NoStop}%
\bibitem [{\citenamefont {Kien}\ \emph {et~al.}(2013)\citenamefont {Kien},
  \citenamefont {Schneeweiss},\ and\ \citenamefont
  {Rauschenbeutel}}]{Kien2013}%
  \BibitemOpen
  \bibfield  {author} {\bibinfo {author} {\bibfnamefont {F.~L.}\ \bibnamefont
  {Kien}}, \bibinfo {author} {\bibfnamefont {P.}~\bibnamefont {Schneeweiss}}, \
  and\ \bibinfo {author} {\bibfnamefont {A.}~\bibnamefont {Rauschenbeutel}},\
  }\href@noop {} {\bibfield  {journal} {\bibinfo  {journal} {Eur. Phys. J. D}\
  }\textbf {\bibinfo {volume} {67}},\ \bibinfo {pages} {1} (\bibinfo {year}
  {2013})}\BibitemShut {NoStop}%
\bibitem [{\citenamefont {{Blundell}}\ \emph {et~al.}(1992)\citenamefont
  {{Blundell}}, \citenamefont {{Sapirstein}},\ and\ \citenamefont
  {{Johnson}}}]{Blundell1992}%
  \BibitemOpen
  \bibfield  {author} {\bibinfo {author} {\bibfnamefont {S.~A.}\ \bibnamefont
  {{Blundell}}}, \bibinfo {author} {\bibfnamefont {J.}~\bibnamefont
  {{Sapirstein}}}, \ and\ \bibinfo {author} {\bibfnamefont {W.~R.}\
  \bibnamefont {{Johnson}}},\ }\href {\doibase 10.1103/PhysRevD.45.1602}
  {\bibfield  {journal} {\bibinfo  {journal} {\prd}\ }\textbf {\bibinfo
  {volume} {45}},\ \bibinfo {pages} {1602} (\bibinfo {year}
  {1992})}\BibitemShut {NoStop}%
\bibitem [{\citenamefont {Patterson}\ \emph {et~al.}(2015)\citenamefont
  {Patterson}, \citenamefont {Sell}, \citenamefont {Ehrenreich}, \citenamefont
  {Gearba}, \citenamefont {Brooke}, \citenamefont {Scoville},\ and\
  \citenamefont {Knize}}]{Patterson2015}%
  \BibitemOpen
  \bibfield  {author} {\bibinfo {author} {\bibfnamefont {B.~M.}\ \bibnamefont
  {Patterson}}, \bibinfo {author} {\bibfnamefont {J.~F.}\ \bibnamefont {Sell}},
  \bibinfo {author} {\bibfnamefont {T.}~\bibnamefont {Ehrenreich}}, \bibinfo
  {author} {\bibfnamefont {M.~A.}\ \bibnamefont {Gearba}}, \bibinfo {author}
  {\bibfnamefont {G.~M.}\ \bibnamefont {Brooke}}, \bibinfo {author}
  {\bibfnamefont {J.}~\bibnamefont {Scoville}}, \ and\ \bibinfo {author}
  {\bibfnamefont {R.~J.}\ \bibnamefont {Knize}},\ }\href {\doibase
  10.1103/PhysRevA.91.012506} {\bibfield  {journal} {\bibinfo  {journal} {Phys.
  Rev. A}\ }\textbf {\bibinfo {volume} {91}},\ \bibinfo {pages} {012506}
  (\bibinfo {year} {2015})}\BibitemShut {NoStop}%
\bibitem [{\citenamefont {Damitz}\ \emph {et~al.}(2019)\citenamefont {Damitz},
  \citenamefont {Toh}, \citenamefont {Putney}, \citenamefont {Tanner},\ and\
  \citenamefont {Elliott}}]{Damitz2019}%
  \BibitemOpen
  \bibfield  {author} {\bibinfo {author} {\bibfnamefont {A.}~\bibnamefont
  {Damitz}}, \bibinfo {author} {\bibfnamefont {G.}~\bibnamefont {Toh}},
  \bibinfo {author} {\bibfnamefont {E.}~\bibnamefont {Putney}}, \bibinfo
  {author} {\bibfnamefont {C.~E.}\ \bibnamefont {Tanner}}, \ and\ \bibinfo
  {author} {\bibfnamefont {D.~S.}\ \bibnamefont {Elliott}},\ }\href {\doibase
  10.1103/PhysRevA.99.062510} {\bibfield  {journal} {\bibinfo  {journal} {Phys.
  Rev. A}\ }\textbf {\bibinfo {volume} {99}},\ \bibinfo {pages} {062510}
  (\bibinfo {year} {2019})}\BibitemShut {NoStop}%
\bibitem [{\citenamefont {{Porsev}}\ \emph {et~al.}(2010)\citenamefont
  {{Porsev}}, \citenamefont {{Beloy}},\ and\ \citenamefont
  {{Derevianko}}}]{Porsev2010}%
  \BibitemOpen
  \bibfield  {author} {\bibinfo {author} {\bibfnamefont {S.~G.}\ \bibnamefont
  {{Porsev}}}, \bibinfo {author} {\bibfnamefont {K.}~\bibnamefont {{Beloy}}}, \
  and\ \bibinfo {author} {\bibfnamefont {A.}~\bibnamefont {{Derevianko}}},\
  }\href {\doibase 10.1103/PhysRevD.82.036008} {\bibfield  {journal} {\bibinfo
  {journal} {\prd}\ }\textbf {\bibinfo {volume} {82}},\ \bibinfo {eid} {036008}
  (\bibinfo {year} {2010})}\BibitemShut {NoStop}%
\bibitem [{\citenamefont {{Shabanova}}\ \emph {et~al.}(1979)\citenamefont
  {{Shabanova}}, \citenamefont {{Monakov}},\ and\ \citenamefont
  {{Khlyustalov}}}]{Shabanova1979}%
  \BibitemOpen
  \bibfield  {author} {\bibinfo {author} {\bibfnamefont {L.~N.}\ \bibnamefont
  {{Shabanova}}}, \bibinfo {author} {\bibfnamefont {Y.~N.}\ \bibnamefont
  {{Monakov}}}, \ and\ \bibinfo {author} {\bibfnamefont {A.~N.}\ \bibnamefont
  {{Khlyustalov}}},\ }\href@noop {} {\bibfield  {journal} {\bibinfo  {journal}
  {Opt. Spectrosc.}\ }\textbf {\bibinfo {volume} {47}},\ \bibinfo {pages} {1}
  (\bibinfo {year} {1979})}\BibitemShut {NoStop}%
\bibitem [{\citenamefont {Blundell}\ \emph {et~al.}(1991)\citenamefont
  {Blundell}, \citenamefont {Johnson},\ and\ \citenamefont
  {Sapirstein}}]{Blundell1991}%
  \BibitemOpen
  \bibfield  {author} {\bibinfo {author} {\bibfnamefont {S.~A.}\ \bibnamefont
  {Blundell}}, \bibinfo {author} {\bibfnamefont {W.~R.}\ \bibnamefont
  {Johnson}}, \ and\ \bibinfo {author} {\bibfnamefont {J.}~\bibnamefont
  {Sapirstein}},\ }\href {\doibase 10.1103/PhysRevA.43.3407} {\bibfield
  {journal} {\bibinfo  {journal} {Phys. Rev. A}\ }\textbf {\bibinfo {volume}
  {43}},\ \bibinfo {pages} {3407} (\bibinfo {year} {1991})}\BibitemShut
  {NoStop}%
\bibitem [{\citenamefont {{Antypas}}\ and\ \citenamefont
  {{Elliott}}(2013)}]{Antypas2013}%
  \BibitemOpen
  \bibfield  {author} {\bibinfo {author} {\bibfnamefont {D.}~\bibnamefont
  {{Antypas}}}\ and\ \bibinfo {author} {\bibfnamefont {D.~S.}\ \bibnamefont
  {{Elliott}}},\ }\href {\doibase 10.1103/PhysRevA.88.052516} {\bibfield
  {journal} {\bibinfo  {journal} {\pra}\ }\textbf {\bibinfo {volume} {88}},\
  \bibinfo {eid} {052516} (\bibinfo {year} {2013})}\BibitemShut {NoStop}%
\bibitem [{\citenamefont {{Vasilyev}}\ \emph {et~al.}(2002)\citenamefont
  {{Vasilyev}}, \citenamefont {{Savukov}}, \citenamefont {{Safronova}},\ and\
  \citenamefont {{Berry}}}]{Vasilyev2002}%
  \BibitemOpen
  \bibfield  {author} {\bibinfo {author} {\bibfnamefont {A.~A.}\ \bibnamefont
  {{Vasilyev}}}, \bibinfo {author} {\bibfnamefont {I.~M.}\ \bibnamefont
  {{Savukov}}}, \bibinfo {author} {\bibfnamefont {M.~S.}\ \bibnamefont
  {{Safronova}}}, \ and\ \bibinfo {author} {\bibfnamefont {H.~G.}\ \bibnamefont
  {{Berry}}},\ }\href {\doibase 10.1103/PhysRevA.66.020101} {\bibfield
  {journal} {\bibinfo  {journal} {\pra}\ }\textbf {\bibinfo {volume} {66}},\
  \bibinfo {eid} {020101} (\bibinfo {year} {2002})}\BibitemShut {NoStop}%
\bibitem [{\citenamefont {Toh}\ \emph {et~al.}(2019)\citenamefont {Toh},
  \citenamefont {Damitz}, \citenamefont {Tanner}, \citenamefont {Johnson},\
  and\ \citenamefont {Elliott}}]{Toh2019}%
  \BibitemOpen
  \bibfield  {author} {\bibinfo {author} {\bibfnamefont {G.}~\bibnamefont
  {Toh}}, \bibinfo {author} {\bibfnamefont {A.}~\bibnamefont {Damitz}},
  \bibinfo {author} {\bibfnamefont {C.~E.}\ \bibnamefont {Tanner}}, \bibinfo
  {author} {\bibfnamefont {W.~R.}\ \bibnamefont {Johnson}}, \ and\ \bibinfo
  {author} {\bibfnamefont {D.~S.}\ \bibnamefont {Elliott}},\ }\href {\doibase
  10.1103/PhysRevLett.123.073002} {\bibfield  {journal} {\bibinfo  {journal}
  {Phys. Rev. Lett.}\ }\textbf {\bibinfo {volume} {123}},\ \bibinfo {pages}
  {073002} (\bibinfo {year} {2019})}\BibitemShut {NoStop}%
\bibitem [{\citenamefont {{Toh}}\ \emph {et~al.}(2019)\citenamefont {{Toh}},
  \citenamefont {{Damitz}}, \citenamefont {{Glotzbach}}, \citenamefont
  {{Quirk}}, \citenamefont {{Stevenson}}, \citenamefont {{Choi}}, \citenamefont
  {{Safronova}},\ and\ \citenamefont {{Elliott}}}]{toh2019A}%
  \BibitemOpen
  \bibfield  {author} {\bibinfo {author} {\bibfnamefont {G.}~\bibnamefont
  {{Toh}}}, \bibinfo {author} {\bibfnamefont {A.}~\bibnamefont {{Damitz}}},
  \bibinfo {author} {\bibfnamefont {N.}~\bibnamefont {{Glotzbach}}}, \bibinfo
  {author} {\bibfnamefont {J.}~\bibnamefont {{Quirk}}}, \bibinfo {author}
  {\bibfnamefont {I.~C.}\ \bibnamefont {{Stevenson}}}, \bibinfo {author}
  {\bibfnamefont {J.}~\bibnamefont {{Choi}}}, \bibinfo {author} {\bibfnamefont
  {M.~S.}\ \bibnamefont {{Safronova}}}, \ and\ \bibinfo {author} {\bibfnamefont
  {D.~S.}\ \bibnamefont {{Elliott}}},\ }\href {\doibase
  10.1103/PhysRevA.99.032504} {\bibfield  {journal} {\bibinfo  {journal}
  {\pra}\ }\textbf {\bibinfo {volume} {99}},\ \bibinfo {eid} {032504} (\bibinfo
  {year} {2019})}\BibitemShut {NoStop}%
\bibitem [{\citenamefont {{Bennett}}\ \emph {et~al.}(1999)\citenamefont
  {{Bennett}}, \citenamefont {{Roberts}},\ and\ \citenamefont
  {{Wieman}}}]{Bennett1999}%
  \BibitemOpen
  \bibfield  {author} {\bibinfo {author} {\bibfnamefont {S.~C.}\ \bibnamefont
  {{Bennett}}}, \bibinfo {author} {\bibfnamefont {J.~L.}\ \bibnamefont
  {{Roberts}}}, \ and\ \bibinfo {author} {\bibfnamefont {C.~E.}\ \bibnamefont
  {{Wieman}}},\ }\href {\doibase 10.1103/PhysRevA.59.R16} {\bibfield  {journal}
  {\bibinfo  {journal} {\pra}\ }\textbf {\bibinfo {volume} {59}},\ \bibinfo
  {pages} {R16} (\bibinfo {year} {1999})}\BibitemShut {NoStop}%
\bibitem [{\citenamefont {{Johnson}}\ \emph {et~al.}(1996)\citenamefont
  {{Johnson}}, \citenamefont {{Liu}},\ and\ \citenamefont
  {{Sapirstein}}}]{Johnson1996}%
  \BibitemOpen
  \bibfield  {author} {\bibinfo {author} {\bibfnamefont {W.~R.}\ \bibnamefont
  {{Johnson}}}, \bibinfo {author} {\bibfnamefont {Z.~W.}\ \bibnamefont
  {{Liu}}}, \ and\ \bibinfo {author} {\bibfnamefont {J.}~\bibnamefont
  {{Sapirstein}}},\ }\href {\doibase 10.1006/adnd.1996.0024} {\bibfield
  {journal} {\bibinfo  {journal} {At.~Data~Nucl.~Data~Tables}\ }\textbf
  {\bibinfo {volume} {64}},\ \bibinfo {pages} {279} (\bibinfo {year}
  {1996})}\BibitemShut {NoStop}%
\bibitem [{\citenamefont {{Johnson}}\ \emph {et~al.}(1987)\citenamefont
  {{Johnson}}, \citenamefont {{Idrees}},\ and\ \citenamefont
  {{Sapirstein}}}]{Johnson1987}%
  \BibitemOpen
  \bibfield  {author} {\bibinfo {author} {\bibfnamefont {W.~R.}\ \bibnamefont
  {{Johnson}}}, \bibinfo {author} {\bibfnamefont {M.}~\bibnamefont {{Idrees}}},
  \ and\ \bibinfo {author} {\bibfnamefont {J.}~\bibnamefont {{Sapirstein}}},\
  }\href {\doibase 10.1103/PhysRevA.35.3218} {\bibfield  {journal} {\bibinfo
  {journal} {\pra}\ }\textbf {\bibinfo {volume} {35}},\ \bibinfo {pages} {3218}
  (\bibinfo {year} {1987})}\BibitemShut {NoStop}%
\bibitem [{\citenamefont {{Young}}\ \emph {et~al.}(1994)\citenamefont
  {{Young}}, \citenamefont {{Hill}}, \citenamefont {{Sibener}}, \citenamefont
  {{Price}}, \citenamefont {{Tanner}}, \citenamefont {{Wieman}},\ and\
  \citenamefont {{Leone}}}]{Young1994}%
  \BibitemOpen
  \bibfield  {author} {\bibinfo {author} {\bibfnamefont {L.}~\bibnamefont
  {{Young}}}, \bibinfo {author} {\bibfnamefont {I.}~\bibnamefont {{Hill}},
  \bibfnamefont {W.~T.}}, \bibinfo {author} {\bibfnamefont {S.~J.}\
  \bibnamefont {{Sibener}}}, \bibinfo {author} {\bibfnamefont {S.~D.}\
  \bibnamefont {{Price}}}, \bibinfo {author} {\bibfnamefont {C.~E.}\
  \bibnamefont {{Tanner}}}, \bibinfo {author} {\bibfnamefont {C.~E.}\
  \bibnamefont {{Wieman}}}, \ and\ \bibinfo {author} {\bibfnamefont {S.~R.}\
  \bibnamefont {{Leone}}},\ }\href {\doibase 10.1103/PhysRevA.50.2174}
  {\bibfield  {journal} {\bibinfo  {journal} {\pra}\ }\textbf {\bibinfo
  {volume} {50}},\ \bibinfo {pages} {2174} (\bibinfo {year}
  {1994})}\BibitemShut {NoStop}%
\bibitem [{\citenamefont {Morton}(2000)}]{Morton2000}%
  \BibitemOpen
  \bibfield  {author} {\bibinfo {author} {\bibfnamefont {D.~C.}\ \bibnamefont
  {Morton}},\ }\href@noop {} {\bibfield  {journal} {\bibinfo  {journal} {The
  Astro phys. J., Suppl. Ser}\ }\textbf {\bibinfo {volume} {130}},\ \bibinfo
  {pages} {403} (\bibinfo {year} {2000})}\BibitemShut {NoStop}%
\bibitem [{\citenamefont {Exton}(1976)}]{EXTON1976309}%
  \BibitemOpen
  \bibfield  {author} {\bibinfo {author} {\bibfnamefont {R.~J.}\ \bibnamefont
  {Exton}},\ }\href@noop {} {\bibfield  {journal} {\bibinfo  {journal} {J.
  Quant. Spectrosc. Radiat. Transfer}\ }\textbf {\bibinfo {volume} {16}},\
  \bibinfo {pages} {309 } (\bibinfo {year} {1976})}\BibitemShut {NoStop}%
\bibitem [{\citenamefont {Pichler}(1976)}]{PICHLER1976}%
  \BibitemOpen
  \bibfield  {author} {\bibinfo {author} {\bibfnamefont {G.}~\bibnamefont
  {Pichler}},\ }\href@noop {} {\bibfield  {journal} {\bibinfo  {journal} {J.
  Quant. Spectrosc. Radiat. Transfer}\ }\textbf {\bibinfo {volume} {16}},\
  \bibinfo {pages} {147 } (\bibinfo {year} {1976})}\BibitemShut {NoStop}%
\bibitem [{\citenamefont {{Rafac}}\ \emph {et~al.}(1994)\citenamefont
  {{Rafac}}, \citenamefont {{Tanner}}, \citenamefont {{Livingston}},
  \citenamefont {{Kukla}}, \citenamefont {{Berry}},\ and\ \citenamefont
  {{Kurtz}}}]{Rafac1994}%
  \BibitemOpen
  \bibfield  {author} {\bibinfo {author} {\bibfnamefont {R.~J.}\ \bibnamefont
  {{Rafac}}}, \bibinfo {author} {\bibfnamefont {C.~E.}\ \bibnamefont
  {{Tanner}}}, \bibinfo {author} {\bibfnamefont {A.~E.}\ \bibnamefont
  {{Livingston}}}, \bibinfo {author} {\bibfnamefont {K.~W.}\ \bibnamefont
  {{Kukla}}}, \bibinfo {author} {\bibfnamefont {H.~G.}\ \bibnamefont
  {{Berry}}}, \ and\ \bibinfo {author} {\bibfnamefont {C.~A.}\ \bibnamefont
  {{Kurtz}}},\ }\href {\doibase 10.1103/PhysRevA.50.R1976} {\bibfield
  {journal} {\bibinfo  {journal} {\pra}\ }\textbf {\bibinfo {volume} {50}},\
  \bibinfo {pages} {R1976} (\bibinfo {year} {1994})}\BibitemShut {NoStop}%
\bibitem [{\citenamefont {{Bouloufa}}\ \emph {et~al.}(2007)\citenamefont
  {{Bouloufa}}, \citenamefont {{Crubellier}},\ and\ \citenamefont
  {{Dulieu}}}]{Bouloufa2007}%
  \BibitemOpen
  \bibfield  {author} {\bibinfo {author} {\bibfnamefont {N.}~\bibnamefont
  {{Bouloufa}}}, \bibinfo {author} {\bibfnamefont {A.}~\bibnamefont
  {{Crubellier}}}, \ and\ \bibinfo {author} {\bibfnamefont {O.}~\bibnamefont
  {{Dulieu}}},\ }\href {\doibase 10.1103/PhysRevA.75.052501} {\bibfield
  {journal} {\bibinfo  {journal} {\pra}\ }\textbf {\bibinfo {volume} {75}},\
  \bibinfo {eid} {052501} (\bibinfo {year} {2007})}\BibitemShut {NoStop}%
\bibitem [{\citenamefont {{Sell}}\ \emph {et~al.}(2011)\citenamefont {{Sell}},
  \citenamefont {{Patterson}}, \citenamefont {{Ehrenreich}}, \citenamefont
  {{Brooke}}, \citenamefont {{Scoville}},\ and\ \citenamefont
  {{Knize}}}]{Sell2011}%
  \BibitemOpen
  \bibfield  {author} {\bibinfo {author} {\bibfnamefont {J.~F.}\ \bibnamefont
  {{Sell}}}, \bibinfo {author} {\bibfnamefont {B.~M.}\ \bibnamefont
  {{Patterson}}}, \bibinfo {author} {\bibfnamefont {T.}~\bibnamefont
  {{Ehrenreich}}}, \bibinfo {author} {\bibfnamefont {G.}~\bibnamefont
  {{Brooke}}}, \bibinfo {author} {\bibfnamefont {J.}~\bibnamefont
  {{Scoville}}}, \ and\ \bibinfo {author} {\bibfnamefont {R.~J.}\ \bibnamefont
  {{Knize}}},\ }\href {\doibase 10.1103/PhysRevA.84.010501} {\bibfield
  {journal} {\bibinfo  {journal} {\pra}\ }\textbf {\bibinfo {volume} {84}},\
  \bibinfo {eid} {010501} (\bibinfo {year} {2011})}\BibitemShut {NoStop}%
\bibitem [{\citenamefont {{Tanner}}\ \emph {et~al.}(1992)\citenamefont
  {{Tanner}}, \citenamefont {{Livingston}}, \citenamefont {{Rafac}},
  \citenamefont {{Serpa}}, \citenamefont {{Kukla}}, \citenamefont {{Berry}},
  \citenamefont {{Young}},\ and\ \citenamefont {{Kurtz}}}]{Livingston1992}%
  \BibitemOpen
  \bibfield  {author} {\bibinfo {author} {\bibfnamefont {C.~E.}\ \bibnamefont
  {{Tanner}}}, \bibinfo {author} {\bibfnamefont {A.~E.}\ \bibnamefont
  {{Livingston}}}, \bibinfo {author} {\bibfnamefont {R.~J.}\ \bibnamefont
  {{Rafac}}}, \bibinfo {author} {\bibfnamefont {F.~G.}\ \bibnamefont
  {{Serpa}}}, \bibinfo {author} {\bibfnamefont {K.~W.}\ \bibnamefont
  {{Kukla}}}, \bibinfo {author} {\bibfnamefont {H.~G.}\ \bibnamefont
  {{Berry}}}, \bibinfo {author} {\bibfnamefont {L.}~\bibnamefont {{Young}}}, \
  and\ \bibinfo {author} {\bibfnamefont {C.~A.}\ \bibnamefont {{Kurtz}}},\
  }\href {\doibase 10.1103/PhysRevLett.69.2765} {\bibfield  {journal} {\bibinfo
   {journal} {\prl}\ }\textbf {\bibinfo {volume} {69}},\ \bibinfo {pages}
  {2765} (\bibinfo {year} {1992})}\BibitemShut {NoStop}%
\bibitem [{\citenamefont {Mitroy}\ \emph {et~al.}(1988)\citenamefont {Mitroy},
  \citenamefont {Griffin}, \citenamefont {Norcross},\ and\ \citenamefont
  {Pindzola}}]{Mitroy1988}%
  \BibitemOpen
  \bibfield  {author} {\bibinfo {author} {\bibfnamefont {J.}~\bibnamefont
  {Mitroy}}, \bibinfo {author} {\bibfnamefont {D.~C.}\ \bibnamefont {Griffin}},
  \bibinfo {author} {\bibfnamefont {D.~W.}\ \bibnamefont {Norcross}}, \ and\
  \bibinfo {author} {\bibfnamefont {M.~S.}\ \bibnamefont {Pindzola}},\ }\href
  {\doibase 10.1103/PhysRevA.38.3339} {\bibfield  {journal} {\bibinfo
  {journal} {Phys. Rev. A}\ }\textbf {\bibinfo {volume} {38}},\ \bibinfo
  {pages} {3339} (\bibinfo {year} {1988})}\BibitemShut {NoStop}%
\bibitem [{\citenamefont {{Marinescu}}\ \emph {et~al.}(1994)\citenamefont
  {{Marinescu}}, \citenamefont {{Sadeghpour}},\ and\ \citenamefont
  {{Dalgarno}}}]{Marinescu1994}%
  \BibitemOpen
  \bibfield  {author} {\bibinfo {author} {\bibfnamefont {M.}~\bibnamefont
  {{Marinescu}}}, \bibinfo {author} {\bibfnamefont {H.~R.}\ \bibnamefont
  {{Sadeghpour}}}, \ and\ \bibinfo {author} {\bibfnamefont {A.}~\bibnamefont
  {{Dalgarno}}},\ }\href {\doibase 10.1103/PhysRevA.49.5103} {\bibfield
  {journal} {\bibinfo  {journal} {\pra}\ }\textbf {\bibinfo {volume} {49}},\
  \bibinfo {pages} {5103} (\bibinfo {year} {1994})}\BibitemShut {NoStop}%
\bibitem [{\citenamefont {Hameed}\ \emph {et~al.}(1968)\citenamefont {Hameed},
  \citenamefont {Herzenberg},\ and\ \citenamefont {James}}]{Hameed1968}%
  \BibitemOpen
  \bibfield  {author} {\bibinfo {author} {\bibfnamefont {S.}~\bibnamefont
  {Hameed}}, \bibinfo {author} {\bibfnamefont {A.}~\bibnamefont {Herzenberg}},
  \ and\ \bibinfo {author} {\bibfnamefont {M.~G.}\ \bibnamefont {James}},\
  }\href@noop {} {\bibfield  {journal} {\bibinfo  {journal} {Journal of Physics
  B Atomic Molecular Physics}\ }\textbf {\bibinfo {volume} {1}},\ \bibinfo
  {pages} {822} (\bibinfo {year} {1968})}\BibitemShut {NoStop}%
\bibitem [{\citenamefont {Caves}\ and\ \citenamefont
  {Dalgarno}(1972)}]{Caves1972}%
  \BibitemOpen
  \bibfield  {author} {\bibinfo {author} {\bibfnamefont {T.~C.}\ \bibnamefont
  {Caves}}\ and\ \bibinfo {author} {\bibfnamefont {A.}~\bibnamefont
  {Dalgarno}},\ }\href@noop {} {\bibfield  {journal} {\bibinfo  {journal} {J.
  Quant. Spectrosc. Radiat. Transfer.}\ }\textbf {\bibinfo {volume} {12}},\
  \bibinfo {pages} {1539} (\bibinfo {year} {1972})}\BibitemShut {NoStop}%
\bibitem [{\citenamefont {Hafner}\ and\ \citenamefont
  {Schwarz}(1978)}]{Hafner1978}%
  \BibitemOpen
  \bibfield  {author} {\bibinfo {author} {\bibfnamefont {P.}~\bibnamefont
  {Hafner}}\ and\ \bibinfo {author} {\bibfnamefont {W.~H.~E.}\ \bibnamefont
  {Schwarz}},\ }\href@noop {} {\bibfield  {journal} {\bibinfo  {journal} {J.
  Phys. B.}\ }\textbf {\bibinfo {volume} {11}},\ \bibinfo {pages} {2975}
  (\bibinfo {year} {1978})}\BibitemShut {NoStop}%
\bibitem [{\citenamefont {{Gregoire}}\ \emph {et~al.}(2015)\citenamefont
  {{Gregoire}}, \citenamefont {{Hromada}}, \citenamefont {{Holmgren}},
  \citenamefont {{Trubko}},\ and\ \citenamefont {{Cronin}}}]{Gregoire2015}%
  \BibitemOpen
  \bibfield  {author} {\bibinfo {author} {\bibfnamefont {M.~D.}\ \bibnamefont
  {{Gregoire}}}, \bibinfo {author} {\bibfnamefont {I.}~\bibnamefont
  {{Hromada}}}, \bibinfo {author} {\bibfnamefont {W.~F.}\ \bibnamefont
  {{Holmgren}}}, \bibinfo {author} {\bibfnamefont {R.}~\bibnamefont
  {{Trubko}}}, \ and\ \bibinfo {author} {\bibfnamefont {A.~D.}\ \bibnamefont
  {{Cronin}}},\ }\href {\doibase 10.1103/PhysRevA.92.052513} {\bibfield
  {journal} {\bibinfo  {journal} {\pra}\ }\textbf {\bibinfo {volume} {92}},\
  \bibinfo {eid} {052513} (\bibinfo {year} {2015})}\BibitemShut {NoStop}%
\bibitem [{\citenamefont {{Arora}}\ \emph {et~al.}(2007)\citenamefont
  {{Arora}}, \citenamefont {{Safronova}},\ and\ \citenamefont
  {{Clark}}}]{Arora2007}%
  \BibitemOpen
  \bibfield  {author} {\bibinfo {author} {\bibfnamefont {B.}~\bibnamefont
  {{Arora}}}, \bibinfo {author} {\bibfnamefont {M.~S.}\ \bibnamefont
  {{Safronova}}}, \ and\ \bibinfo {author} {\bibfnamefont {C.~W.}\ \bibnamefont
  {{Clark}}},\ }\href {\doibase 10.1103/PhysRevA.76.052516} {\bibfield
  {journal} {\bibinfo  {journal} {\pra}\ }\textbf {\bibinfo {volume} {76}},\
  \bibinfo {eid} {052516} (\bibinfo {year} {2007})}\BibitemShut {NoStop}%
\bibitem [{\citenamefont {{Dzuba}}\ \emph {et~al.}(1989)\citenamefont
  {{Dzuba}}, \citenamefont {{Flambaum}},\ and\ \citenamefont
  {{Sushkov}}}]{Dzuba1989}%
  \BibitemOpen
  \bibfield  {author} {\bibinfo {author} {\bibfnamefont {V.~A.}\ \bibnamefont
  {{Dzuba}}}, \bibinfo {author} {\bibfnamefont {V.~V.}\ \bibnamefont
  {{Flambaum}}}, \ and\ \bibinfo {author} {\bibfnamefont {O.~P.}\ \bibnamefont
  {{Sushkov}}},\ }\href {\doibase 10.1016/0375-9601(89)90777-9} {\bibfield
  {journal} {\bibinfo  {journal} {Phys. Lett. A}\ }\textbf {\bibinfo {volume}
  {141}},\ \bibinfo {pages} {147} (\bibinfo {year} {1989})}\BibitemShut
  {NoStop}%
\bibitem [{\citenamefont {Mitroy}\ \emph {et~al.}(2010)\citenamefont {Mitroy},
  \citenamefont {Safronova},\ and\ \citenamefont {Clark}}]{mitroy2010}%
  \BibitemOpen
  \bibfield  {author} {\bibinfo {author} {\bibfnamefont {J.}~\bibnamefont
  {Mitroy}}, \bibinfo {author} {\bibfnamefont {M.~S.}\ \bibnamefont
  {Safronova}}, \ and\ \bibinfo {author} {\bibfnamefont {C.~W.}\ \bibnamefont
  {Clark}},\ }\href@noop {} {\bibfield  {journal} {\bibinfo  {journal} {J.
  Phys. B}\ }\textbf {\bibinfo {volume} {43}},\ \bibinfo {pages} {202001}
  (\bibinfo {year} {2010})}\BibitemShut {NoStop}%
\bibitem [{\citenamefont {{Derevianko}}\ \emph {et~al.}(1999)\citenamefont
  {{Derevianko}}, \citenamefont {{Johnson}}, \citenamefont {{Safronova}},\ and\
  \citenamefont {{Babb}}}]{Derevianko1999}%
  \BibitemOpen
  \bibfield  {author} {\bibinfo {author} {\bibfnamefont {A.}~\bibnamefont
  {{Derevianko}}}, \bibinfo {author} {\bibfnamefont {W.~R.}\ \bibnamefont
  {{Johnson}}}, \bibinfo {author} {\bibfnamefont {M.~S.}\ \bibnamefont
  {{Safronova}}}, \ and\ \bibinfo {author} {\bibfnamefont {J.~F.}\ \bibnamefont
  {{Babb}}},\ }\href {\doibase 10.1103/PhysRevLett.82.3589} {\bibfield
  {journal} {\bibinfo  {journal} {\prl}\ }\textbf {\bibinfo {volume} {82}},\
  \bibinfo {pages} {3589} (\bibinfo {year} {1999})}\BibitemShut {NoStop}%
\bibitem [{\citenamefont {{Singh}}\ \emph {et~al.}(2016)\citenamefont
  {{Singh}}, \citenamefont {{Kaur}}, \citenamefont {{Sahoo}},\ and\
  \citenamefont {{Arora}}}]{Singh2016}%
  \BibitemOpen
  \bibfield  {author} {\bibinfo {author} {\bibfnamefont {S.}~\bibnamefont
  {{Singh}}}, \bibinfo {author} {\bibfnamefont {K.}~\bibnamefont {{Kaur}}},
  \bibinfo {author} {\bibfnamefont {B.~K.}\ \bibnamefont {{Sahoo}}}, \ and\
  \bibinfo {author} {\bibfnamefont {B.}~\bibnamefont {{Arora}}},\ }\href
  {\doibase 10.1088/0953-4075/49/14/145005} {\bibfield  {journal} {\bibinfo
  {journal} {J. Phys. B}\ }\textbf {\bibinfo {volume} {49}},\ \bibinfo {eid}
  {145005} (\bibinfo {year} {2016})}\BibitemShut {NoStop}%
\bibitem [{\citenamefont {{Iskrenova-Tchoukova}}\ \emph
  {et~al.}(2007)\citenamefont {{Iskrenova-Tchoukova}}, \citenamefont
  {{Safronova}},\ and\ \citenamefont {{Safronova}}}]{Iskrenova2007}%
  \BibitemOpen
  \bibfield  {author} {\bibinfo {author} {\bibfnamefont {E.}~\bibnamefont
  {{Iskrenova-Tchoukova}}}, \bibinfo {author} {\bibfnamefont {M.~S.}\
  \bibnamefont {{Safronova}}}, \ and\ \bibinfo {author} {\bibfnamefont {U.~I.}\
  \bibnamefont {{Safronova}}},\ }\href {\doibase 10.3233/JCM-2007-75-616}
  {\bibfield  {journal} {\bibinfo  {journal} {J. Comp. Methods in Sci. and
  Eng.}\ }\textbf {\bibinfo {volume} {7}},\ \bibinfo {pages} {521} (\bibinfo
  {year} {2007})}\BibitemShut {NoStop}%
\bibitem [{\citenamefont {{Lim}}\ \emph {et~al.}(2005)\citenamefont {{Lim}},
  \citenamefont {{Schwerdtfeger}}, \citenamefont {{Metz}},\ and\ \citenamefont
  {{Stoll}}}]{Lim2005}%
  \BibitemOpen
  \bibfield  {author} {\bibinfo {author} {\bibfnamefont {I.~S.}\ \bibnamefont
  {{Lim}}}, \bibinfo {author} {\bibfnamefont {P.}~\bibnamefont
  {{Schwerdtfeger}}}, \bibinfo {author} {\bibfnamefont {B.}~\bibnamefont
  {{Metz}}}, \ and\ \bibinfo {author} {\bibfnamefont {H.}~\bibnamefont
  {{Stoll}}},\ }\href {\doibase 10.1063/1.1856451} {\bibfield  {journal}
  {\bibinfo  {journal} {\jcp}\ }\textbf {\bibinfo {volume} {122}},\ \bibinfo
  {pages} {104103} (\bibinfo {year} {2005})}\BibitemShut {NoStop}%
\bibitem [{\citenamefont {{Borschevsky}}\ \emph {et~al.}(2013)\citenamefont
  {{Borschevsky}}, \citenamefont {{Pershina}}, \citenamefont {{Eliav}},\ and\
  \citenamefont {{Kaldor}}}]{Borschevsky2013}%
  \BibitemOpen
  \bibfield  {author} {\bibinfo {author} {\bibfnamefont {A.}~\bibnamefont
  {{Borschevsky}}}, \bibinfo {author} {\bibfnamefont {V.}~\bibnamefont
  {{Pershina}}}, \bibinfo {author} {\bibfnamefont {E.}~\bibnamefont {{Eliav}}},
  \ and\ \bibinfo {author} {\bibfnamefont {U.}~\bibnamefont {{Kaldor}}},\
  }\href {\doibase 10.1063/1.4795433} {\bibfield  {journal} {\bibinfo
  {journal} {\jcp}\ }\textbf {\bibinfo {volume} {138}},\ \bibinfo {pages}
  {124302} (\bibinfo {year} {2013})}\BibitemShut {NoStop}%
\bibitem [{\citenamefont {{Lim}}\ \emph {et~al.}(1999)\citenamefont {{Lim}},
  \citenamefont {{Pernpointner}}, \citenamefont {{Seth}}, \citenamefont
  {{Laerdahl}}, \citenamefont {{Schwerdtfeger}}, \citenamefont {{Neogrady}},\
  and\ \citenamefont {{Urban}}}]{Lim1999}%
  \BibitemOpen
  \bibfield  {author} {\bibinfo {author} {\bibfnamefont {I.~S.}\ \bibnamefont
  {{Lim}}}, \bibinfo {author} {\bibfnamefont {M.}~\bibnamefont
  {{Pernpointner}}}, \bibinfo {author} {\bibfnamefont {M.}~\bibnamefont
  {{Seth}}}, \bibinfo {author} {\bibfnamefont {J.~K.}\ \bibnamefont
  {{Laerdahl}}}, \bibinfo {author} {\bibfnamefont {P.}~\bibnamefont
  {{Schwerdtfeger}}}, \bibinfo {author} {\bibfnamefont {P.}~\bibnamefont
  {{Neogrady}}}, \ and\ \bibinfo {author} {\bibfnamefont {M.}~\bibnamefont
  {{Urban}}},\ }\href {\doibase 10.1103/PhysRevA.60.2822} {\bibfield  {journal}
  {\bibinfo  {journal} {\pra}\ }\textbf {\bibinfo {volume} {60}},\ \bibinfo
  {pages} {2822} (\bibinfo {year} {1999})}\BibitemShut {NoStop}%
\bibitem [{\citenamefont {{Amini}}\ and\ \citenamefont
  {{Gould}}(2003)}]{Amini2003}%
  \BibitemOpen
  \bibfield  {author} {\bibinfo {author} {\bibfnamefont {J.~M.}\ \bibnamefont
  {{Amini}}}\ and\ \bibinfo {author} {\bibfnamefont {H.}~\bibnamefont
  {{Gould}}},\ }\href {\doibase 10.1103/PhysRevLett.91.153001} {\bibfield
  {journal} {\bibinfo  {journal} {\prl}\ }\textbf {\bibinfo {volume} {91}},\
  \bibinfo {eid} {153001} (\bibinfo {year} {2003})}\BibitemShut {NoStop}%
\bibitem [{Sup(CICP)}]{Suppl}%
  \BibitemOpen
  \href@noop {} {}\ (\bibinfo {year} {See Supplemental Material at
  http://XXXXXX for additional resutls of RCICP.})\BibitemShut {NoStop}%
\bibitem [{\citenamefont {Williams}\ \emph {et~al.}(2018)\citenamefont
  {Williams}, \citenamefont {Herd},\ and\ \citenamefont
  {Hawkins}}]{Williams2018}%
  \BibitemOpen
  \bibfield  {author} {\bibinfo {author} {\bibfnamefont {W.~D.}\ \bibnamefont
  {Williams}}, \bibinfo {author} {\bibfnamefont {M.~T.}\ \bibnamefont {Herd}},
  \ and\ \bibinfo {author} {\bibfnamefont {W.~B.}\ \bibnamefont {Hawkins}},\
  }\href {\doibase 10.1088/1612-202x/aac97e} {\bibfield  {journal} {\bibinfo
  {journal} {Laser. Phys. Lett.}\ }\textbf {\bibinfo {volume} {15}},\ \bibinfo
  {pages} {095702} (\bibinfo {year} {2018})}\BibitemShut {NoStop}%
\bibitem [{\citenamefont {Carr}\ and\ \citenamefont
  {Saffman}(2016)}]{Carr2016}%
  \BibitemOpen
  \bibfield  {author} {\bibinfo {author} {\bibfnamefont {A.~W.}\ \bibnamefont
  {Carr}}\ and\ \bibinfo {author} {\bibfnamefont {M.}~\bibnamefont {Saffman}},\
  }\href {\doibase 10.1103/PhysRevLett.117.150801} {\bibfield  {journal}
  {\bibinfo  {journal} {Phys. Rev. Lett.}\ }\textbf {\bibinfo {volume} {117}},\
  \bibinfo {pages} {150801} (\bibinfo {year} {2016})}\BibitemShut {NoStop}%
\bibitem [{\citenamefont {Gerginov}\ \emph {et~al.}(2006)\citenamefont
  {Gerginov}, \citenamefont {Calkins}, \citenamefont {Tanner}, \citenamefont
  {McFerran}, \citenamefont {Diddams}, \citenamefont {Bartels},\ and\
  \citenamefont {Hollberg}}]{Gerginov2006}%
  \BibitemOpen
  \bibfield  {author} {\bibinfo {author} {\bibfnamefont {V.}~\bibnamefont
  {Gerginov}}, \bibinfo {author} {\bibfnamefont {K.}~\bibnamefont {Calkins}},
  \bibinfo {author} {\bibfnamefont {C.~E.}\ \bibnamefont {Tanner}}, \bibinfo
  {author} {\bibfnamefont {J.~J.}\ \bibnamefont {McFerran}}, \bibinfo {author}
  {\bibfnamefont {S.}~\bibnamefont {Diddams}}, \bibinfo {author} {\bibfnamefont
  {A.}~\bibnamefont {Bartels}}, \ and\ \bibinfo {author} {\bibfnamefont
  {L.}~\bibnamefont {Hollberg}},\ }\href {\doibase 10.1103/PhysRevA.73.032504}
  {\bibfield  {journal} {\bibinfo  {journal} {Phys. Rev. A}\ }\textbf {\bibinfo
  {volume} {73}},\ \bibinfo {pages} {032504} (\bibinfo {year}
  {2006})}\BibitemShut {NoStop}%
\bibitem [{\citenamefont {Johnson}\ \emph {et~al.}(2004)\citenamefont
  {Johnson}, \citenamefont {Ho}, \citenamefont {Tanner},\ and\ \citenamefont
  {Derevianko}}]{Johnson2004}%
  \BibitemOpen
  \bibfield  {author} {\bibinfo {author} {\bibfnamefont {W.~R.}\ \bibnamefont
  {Johnson}}, \bibinfo {author} {\bibfnamefont {H.~C.}\ \bibnamefont {Ho}},
  \bibinfo {author} {\bibfnamefont {C.~E.}\ \bibnamefont {Tanner}}, \ and\
  \bibinfo {author} {\bibfnamefont {A.}~\bibnamefont {Derevianko}},\ }\href
  {\doibase 10.1103/PhysRevA.70.014501} {\bibfield  {journal} {\bibinfo
  {journal} {Phys. Rev. A}\ }\textbf {\bibinfo {volume} {70}},\ \bibinfo
  {pages} {014501} (\bibinfo {year} {2004})}\BibitemShut {NoStop}%
\bibitem [{\citenamefont {Feiertag}\ \emph {et~al.}(1972)\citenamefont
  {Feiertag}, \citenamefont {Sahm},\ and\ \citenamefont
  {Putlitz}}]{Feiertag1972Core}%
  \BibitemOpen
  \bibfield  {author} {\bibinfo {author} {\bibfnamefont {D.}~\bibnamefont
  {Feiertag}}, \bibinfo {author} {\bibfnamefont {A.}~\bibnamefont {Sahm}}, \
  and\ \bibinfo {author} {\bibfnamefont {G.~Z.}\ \bibnamefont {Putlitz}},\
  }\href@noop {} {\bibfield  {journal} {\bibinfo  {journal} {Zeitschrift Für
  Physik A Hadrons Nuclei}\ }\textbf {\bibinfo {volume} {255}},\ \bibinfo
  {pages} {93} (\bibinfo {year} {1972})}\BibitemShut {NoStop}%
\bibitem [{\citenamefont {Belin}\ \emph {et~al.}(1976)\citenamefont {Belin},
  \citenamefont {Holmgren},\ and\ \citenamefont {Svanberg}}]{Belin1976}%
  \BibitemOpen
  \bibfield  {author} {\bibinfo {author} {\bibfnamefont {G.}~\bibnamefont
  {Belin}}, \bibinfo {author} {\bibfnamefont {L.}~\bibnamefont {Holmgren}}, \
  and\ \bibinfo {author} {\bibfnamefont {S.}~\bibnamefont {Svanberg}},\
  }\href@noop {} {\bibfield  {journal} {\bibinfo  {journal} {Phys. Scr.}\
  }\textbf {\bibinfo {volume} {14}},\ \bibinfo {pages} {39} (\bibinfo {year}
  {1976})}\BibitemShut {NoStop}%
\bibitem [{\citenamefont {S.}\ \emph {et~al.}(1969)\citenamefont {S.},
  \citenamefont {Svanberg}, \citenamefont {S.},\ and\ \citenamefont
  {Rydberg}}]{S2005Level}%
  \BibitemOpen
  \bibfield  {author} {\bibinfo {author} {\bibnamefont {S.}}, \bibinfo {author}
  {\bibnamefont {Svanberg}}, \bibinfo {author} {\bibnamefont {S.}}, \ and\
  \bibinfo {author} {\bibnamefont {Rydberg}},\ }\href@noop {} {\bibfield
  {journal} {\bibinfo  {journal} {Zeitschrift F{\"u}r Physik A Hadrons Nuclei}\
  }\textbf {\bibinfo {volume} {227}},\ \bibinfo {pages} {216} (\bibinfo {year}
  {1969})}\BibitemShut {NoStop}%
\bibitem [{\citenamefont {{Feichtner}}\ \emph {et~al.}(1965)\citenamefont
  {{Feichtner}}, \citenamefont {{Hoover}},\ and\ \citenamefont
  {{Mizushima}}}]{Feichtner1965}%
  \BibitemOpen
  \bibfield  {author} {\bibinfo {author} {\bibfnamefont {J.~D.}\ \bibnamefont
  {{Feichtner}}}, \bibinfo {author} {\bibfnamefont {M.~E.}\ \bibnamefont
  {{Hoover}}}, \ and\ \bibinfo {author} {\bibfnamefont {M.}~\bibnamefont
  {{Mizushima}}},\ }\href {\doibase 10.1103/PhysRev.137.A702} {\bibfield
  {journal} {\bibinfo  {journal} {Phys. Rev.}\ }\textbf {\bibinfo {volume}
  {137}},\ \bibinfo {pages} {702} (\bibinfo {year} {1965})}\BibitemShut
  {NoStop}%
\bibitem [{\citenamefont {Lee}\ \emph {et~al.}(1975)\citenamefont {Lee},
  \citenamefont {Das},\ and\ \citenamefont {Sternheimer}}]{Lee1975}%
  \BibitemOpen
  \bibfield  {author} {\bibinfo {author} {\bibfnamefont {T.}~\bibnamefont
  {Lee}}, \bibinfo {author} {\bibfnamefont {T.~P.}\ \bibnamefont {Das}}, \ and\
  \bibinfo {author} {\bibfnamefont {R.~M.}\ \bibnamefont {Sternheimer}},\
  }\href {\doibase 10.1103/PhysRevA.11.1784} {\bibfield  {journal} {\bibinfo
  {journal} {Phys. Rev. A}\ }\textbf {\bibinfo {volume} {11}},\ \bibinfo
  {pages} {1784} (\bibinfo {year} {1975})}\BibitemShut {NoStop}%
\bibitem [{\citenamefont {{Pal'chikov}}\ \emph {et~al.}(2003)\citenamefont
  {{Pal'chikov}}, \citenamefont {{Domnin}},\ and\ \citenamefont
  {{Novoselov}}}]{Palchikov2003}%
  \BibitemOpen
  \bibfield  {author} {\bibinfo {author} {\bibfnamefont {V.~G.}\ \bibnamefont
  {{Pal'chikov}}}, \bibinfo {author} {\bibfnamefont {Y.~S.}\ \bibnamefont
  {{Domnin}}}, \ and\ \bibinfo {author} {\bibfnamefont {A.~V.}\ \bibnamefont
  {{Novoselov}}},\ }\href@noop {} {\bibfield  {journal} {\bibinfo  {journal}
  {J. Opt. B}\ }\textbf {\bibinfo {volume} {5}},\ \bibinfo {pages} {S131}
  (\bibinfo {year} {2003})}\BibitemShut {NoStop}%
\bibitem [{\citenamefont {{Micalizio}}\ \emph {et~al.}(2004)\citenamefont
  {{Micalizio}}, \citenamefont {{Godone}}, \citenamefont {{Calonico}},
  \citenamefont {{Levi}},\ and\ \citenamefont {{Lorini}}}]{Micalizio2004}%
  \BibitemOpen
  \bibfield  {author} {\bibinfo {author} {\bibfnamefont {S.}~\bibnamefont
  {{Micalizio}}}, \bibinfo {author} {\bibfnamefont {A.}~\bibnamefont
  {{Godone}}}, \bibinfo {author} {\bibfnamefont {D.}~\bibnamefont
  {{Calonico}}}, \bibinfo {author} {\bibfnamefont {F.}~\bibnamefont {{Levi}}},
  \ and\ \bibinfo {author} {\bibfnamefont {L.}~\bibnamefont {{Lorini}}},\
  }\href {\doibase 10.1103/PhysRevA.69.053401} {\bibfield  {journal} {\bibinfo
  {journal} {\pra}\ }\textbf {\bibinfo {volume} {69}},\ \bibinfo {eid} {053401}
  (\bibinfo {year} {2004})}\BibitemShut {NoStop}%
\bibitem [{\citenamefont {{Angstmann}}\ \emph {et~al.}(2006)\citenamefont
  {{Angstmann}}, \citenamefont {{Dzuba}},\ and\ \citenamefont
  {{Flambaum}}}]{Angstmann2006}%
  \BibitemOpen
  \bibfield  {author} {\bibinfo {author} {\bibfnamefont {E.~J.}\ \bibnamefont
  {{Angstmann}}}, \bibinfo {author} {\bibfnamefont {V.~A.}\ \bibnamefont
  {{Dzuba}}}, \ and\ \bibinfo {author} {\bibfnamefont {V.~V.}\ \bibnamefont
  {{Flambaum}}},\ }\href {\doibase 10.1103/PhysRevA.74.023405} {\bibfield
  {journal} {\bibinfo  {journal} {\pra}\ }\textbf {\bibinfo {volume} {74}},\
  \bibinfo {eid} {023405} (\bibinfo {year} {2006})}\BibitemShut {NoStop}%
\bibitem [{\citenamefont {{Haun}}\ and\ \citenamefont
  {{Zacharias}}(1957)}]{Haun1957}%
  \BibitemOpen
  \bibfield  {author} {\bibinfo {author} {\bibfnamefont {R.~D.}\ \bibnamefont
  {{Haun}}}\ and\ \bibinfo {author} {\bibfnamefont {J.~R.}\ \bibnamefont
  {{Zacharias}}},\ }\href {\doibase 10.1103/PhysRev.107.107} {\bibfield
  {journal} {\bibinfo  {journal} {Phys. Rev.}\ }\textbf {\bibinfo {volume}
  {107}},\ \bibinfo {pages} {107} (\bibinfo {year} {1957})}\BibitemShut
  {NoStop}%
\bibitem [{\citenamefont {{Mowat}}(1972)}]{Mowat1972}%
  \BibitemOpen
  \bibfield  {author} {\bibinfo {author} {\bibfnamefont {J.~R.}\ \bibnamefont
  {{Mowat}}},\ }\href {\doibase 10.1103/PhysRevA.5.1059} {\bibfield  {journal}
  {\bibinfo  {journal} {\pra}\ }\textbf {\bibinfo {volume} {5}},\ \bibinfo
  {pages} {1059} (\bibinfo {year} {1972})}\BibitemShut {NoStop}%
\bibitem [{\citenamefont {{Bauch}}\ and\ \citenamefont
  {{Schr{\"o}der}}(1997)}]{Bauch1997}%
  \BibitemOpen
  \bibfield  {author} {\bibinfo {author} {\bibfnamefont {A.}~\bibnamefont
  {{Bauch}}}\ and\ \bibinfo {author} {\bibfnamefont {R.}~\bibnamefont
  {{Schr{\"o}der}}},\ }\href {\doibase 10.1103/PhysRevLett.78.622} {\bibfield
  {journal} {\bibinfo  {journal} {\prl}\ }\textbf {\bibinfo {volume} {78}},\
  \bibinfo {pages} {622} (\bibinfo {year} {1997})}\BibitemShut {NoStop}%
\bibitem [{\citenamefont {{Simon}}\ \emph {et~al.}(1998)\citenamefont
  {{Simon}}, \citenamefont {{Laurent}},\ and\ \citenamefont
  {{Clairon}}}]{Simon1998}%
  \BibitemOpen
  \bibfield  {author} {\bibinfo {author} {\bibfnamefont {E.}~\bibnamefont
  {{Simon}}}, \bibinfo {author} {\bibfnamefont {P.}~\bibnamefont {{Laurent}}},
  \ and\ \bibinfo {author} {\bibfnamefont {A.}~\bibnamefont {{Clairon}}},\
  }\href {\doibase 10.1103/PhysRevA.57.436} {\bibfield  {journal} {\bibinfo
  {journal} {\pra}\ }\textbf {\bibinfo {volume} {57}},\ \bibinfo {pages} {436}
  (\bibinfo {year} {1998})}\BibitemShut {NoStop}%
\bibitem [{\citenamefont {{Levi}}\ \emph {et~al.}(2004)\citenamefont {{Levi}},
  \citenamefont {{Calonico}}, \citenamefont {{Lorini}}, \citenamefont
  {{Micalizio}},\ and\ \citenamefont {{Godone}}}]{Levi2004}%
  \BibitemOpen
  \bibfield  {author} {\bibinfo {author} {\bibfnamefont {F.}~\bibnamefont
  {{Levi}}}, \bibinfo {author} {\bibfnamefont {D.}~\bibnamefont {{Calonico}}},
  \bibinfo {author} {\bibfnamefont {L.}~\bibnamefont {{Lorini}}}, \bibinfo
  {author} {\bibfnamefont {S.}~\bibnamefont {{Micalizio}}}, \ and\ \bibinfo
  {author} {\bibfnamefont {A.}~\bibnamefont {{Godone}}},\ }\href {\doibase
  10.1103/PhysRevA.70.033412} {\bibfield  {journal} {\bibinfo  {journal}
  {\pra}\ }\textbf {\bibinfo {volume} {70}},\ \bibinfo {eid} {033412} (\bibinfo
  {year} {2004})}\BibitemShut {NoStop}%
\bibitem [{\citenamefont {{Godone}}\ \emph {et~al.}(2005)\citenamefont
  {{Godone}}, \citenamefont {{Calonico}}, \citenamefont {{Levi}}, \citenamefont
  {{Micalizio}},\ and\ \citenamefont {{Calosso}}}]{Godone2005}%
  \BibitemOpen
  \bibfield  {author} {\bibinfo {author} {\bibfnamefont {A.}~\bibnamefont
  {{Godone}}}, \bibinfo {author} {\bibfnamefont {D.}~\bibnamefont
  {{Calonico}}}, \bibinfo {author} {\bibfnamefont {F.}~\bibnamefont {{Levi}}},
  \bibinfo {author} {\bibfnamefont {S.}~\bibnamefont {{Micalizio}}}, \ and\
  \bibinfo {author} {\bibfnamefont {C.}~\bibnamefont {{Calosso}}},\ }\href
  {\doibase 10.1103/PhysRevA.71.063401} {\bibfield  {journal} {\bibinfo
  {journal} {\pra}\ }\textbf {\bibinfo {volume} {71}},\ \bibinfo {eid} {063401}
  (\bibinfo {year} {2005})}\BibitemShut {NoStop}%
\bibitem [{\citenamefont {{Beloy}}\ \emph {et~al.}(2006)\citenamefont
  {{Beloy}}, \citenamefont {{Safronova}},\ and\ \citenamefont
  {{Derevianko}}}]{Beloy2006}%
  \BibitemOpen
  \bibfield  {author} {\bibinfo {author} {\bibfnamefont {K.}~\bibnamefont
  {{Beloy}}}, \bibinfo {author} {\bibfnamefont {U.~I.}\ \bibnamefont
  {{Safronova}}}, \ and\ \bibinfo {author} {\bibfnamefont {A.}~\bibnamefont
  {{Derevianko}}},\ }\href {\doibase 10.1103/PhysRevLett.97.040801} {\bibfield
  {journal} {\bibinfo  {journal} {\prl}\ }\textbf {\bibinfo {volume} {97}},\
  \bibinfo {eid} {040801} (\bibinfo {year} {2006})}\BibitemShut {NoStop}%
\bibitem [{\citenamefont {Ulzega}\ \emph {et~al.}(2006)\citenamefont {Ulzega},
  \citenamefont {Hofer}, \citenamefont {Moroshkin},\ and\ \citenamefont
  {Weis}}]{2006Stark}%
  \BibitemOpen
  \bibfield  {author} {\bibinfo {author} {\bibfnamefont {S.}~\bibnamefont
  {Ulzega}}, \bibinfo {author} {\bibfnamefont {A.}~\bibnamefont {Hofer}},
  \bibinfo {author} {\bibfnamefont {P.}~\bibnamefont {Moroshkin}}, \ and\
  \bibinfo {author} {\bibfnamefont {A.}~\bibnamefont {Weis}},\ }\href@noop {}
  {\bibfield  {journal} {\bibinfo  {journal} {arXive preprint
  physics/0604233.}\ } (\bibinfo {year} {2006})}\BibitemShut {NoStop}%
\bibitem [{\citenamefont {Kramida}(2013)}]{kramida2013c}%
  \BibitemOpen
  \bibfield  {author} {\bibinfo {author} {\bibfnamefont {A.}~\bibnamefont
  {Kramida}},\ }\href@noop {} {\bibfield  {journal} {\bibinfo  {journal}
  {Fusion. Sci. Technol.}\ }\textbf {\bibinfo {volume} {63}},\ \bibinfo {pages}
  {313} (\bibinfo {year} {2013})}\BibitemShut {NoStop}%
\end{thebibliography}

%


%

\end{document}